\documentclass[aps,prb,twocolumn,superscriptaddress,showpacs]{revtex4}
\usepackage{graphicx}

\usepackage{amsmath}
\usepackage{booktabs}
\usepackage{color}
\usepackage{epstopdf}
\usepackage[colorlinks=true, urlcolor=blue, citecolor=blue, linkcolor=blue]{hyperref}

\begin{document}

\title{Superadiabatic quantum heat engine with a multiferroic  working medium}
\author{L. Chotorlishvili}
\affiliation{Institut f\"ur Physik, Martin-Luther-Universit\"at Halle-Wittenberg, 06099 Halle, Germany}
\author{M. Azimi}
\affiliation{Institut f\"ur Physik, Martin-Luther-Universit\"at Halle-Wittenberg, 06099 Halle, Germany}
\author{S. Stagraczy\'nski}
\affiliation{Institut f\"ur Physik, Martin-Luther-Universit\"at Halle-Wittenberg, 06099 Halle, Germany}
\author{Z. Toklikishvili}
\affiliation{Department of Physics, Tbilisi State University, Chavchavadze av. 3, 0128, Tbilisi, Georgia}
\author{M. Sch\"uler}
\affiliation{Institut f\"ur Physik, Martin-Luther-Universit\"at Halle-Wittenberg, 06099 Halle, Germany}
\author{J. Berakdar}
\affiliation{Institut f\"ur Physik, Martin-Luther-Universit\"at Halle-Wittenberg, 06099 Halle, Germany}

\begin{abstract}

A quantum thermodynamic cycle with a chiral multiferroic working substance such as $\textrm{LiCu}_{2}\textrm{O}_{2}$ is presented.
Shortcuts to adiabaticity are employed to achieve an efficient, finite time quantum thermodynamic cycle which is found to depend on the spin ordering.  The emergent electric polarization associated with the  chiral spin order, i.e. the magnetoelectric coupling,  renders possible steering of the spin order by an external electric field and hence renders possible  an electric-field control of the cycle.
Due to the intrinsic coupling between of the spin and the electric polarization, the
cycle performs  an electro-magnetic work.
We determine this work's mean square fluctuations, the irreversible work, and the output power of the cycle. We observe that the work mean square fluctuations are increased with the duration of the adiabatic strokes while the irreversible work and the output power of the cycle show a non-monotonic behavior. In particular the irreversible work vanishes at the end of the quantum adiabatic
strokes. This fact confirms that the cycle is reversible. Our theoretical findings evidence the existence of a system inherent maximal output power. By implementing a Lindblad master equation we  quantify  the role of thermal relaxations on the cycle efficiency.
We also discuss the role of entanglement encoded in the  non-collinear  spin order  as a resource to
affect the quantum thermodynamic cycle.

\end{abstract}
\date{\today}

\maketitle

\section{Introduction}
Analyzing the correspondence and the transition between the quantum and classical regimes is of a fundamental interest as well as useful for understanding physical
processes \cite{Landau1}. Connections  between classical (statistical) mechanics  and quantum statistics are well established \cite{Landau1,Landau2}.
Concerning the quantum transitions of thermodynamic properties,   the situation is  intricate \cite{Campisi}. For mesoscopic quantum systems not only the
size but also quantumness are important  for fluctuation relations.
In particular, for  heat engines operating with a quantum working substance \cite{Altintas,Esposito2010, Georgescu,Rossnagel,Jarzynski1,delCampo2014,delCampo,Wang2,Esposito,Poletti}, the concept of adiabaticity and thermalization should be revisited:

Perturbing externally the Hamiltonian of the working substance  leads to  inter-level transitions of purely quantum origin.  One can exclude thermally assisted inter-level transitions by detaching from the heat bath. For quantum adiabaticity this is however not enough.
   Due to  pure quantum inter-level transitions, the stroke of the cycle which is adiabatic in the classical thermodynamic sense
    may be non-adiabatic for a quantum working substance.
    Quantum adiabaticity implies not only  a decoupling of the system from the  thermal source but also requires an elimination of  inter-level transitions that are of a pure quantum nature. Aside from this aspect, a desirable  feature of a (quantum) heat engine is not  a high efficiency and a slow cycle but rather a good  efficiency at maximal power. To this end, a quantum thermodynamic cycle should be carried out within a finite time.
    To this end, the concept of  shortcuts to quantum adiabaticity is useful \cite{Demi,Berry}.

 This technique quenches the effect of inter-level transitions that are of a pure quantum origin. Such transitions naturally accompany fast driving processes. Nevertheless, shortcuts to quantum adiabaticity eliminate the effect of those inter-level transitions allowing for a quantum heat engine with a finite output power.
    A further central point is the appropriate working substance.
      We identified  multiferroics (MF) and in particular magnetoelectrics nanostructures as promising candidates\cite{Azimi1,Azimi2}.
       MFs possess intrinsically  coupled order parameters such as elastic, magnetic, and ferroelectric orders   \cite{Wang3,Bibes,Dawber,Valencia,Sahoo,Menzel,Mostovoy,rep_prog_phys,fiebig_science} and can be well integrated in electronic circuites (in particular in oxide-based electronics). Hence, an engine based on a MF substance performs  magnetic, electric and possibly (via piezoelectricity) mechanical works, at the same time.
       A particularly interesting case is that of a quantum spiral magnetoelectric substance\cite{Azimi1,Azimi2}.

We note that the employed model applies to  experimentally feasible systems. Experiments on ferroelectricity and magnetoelectric coupling in the spiral-magnetic state of the 1D quantum magnet LiCu$_{2}$O$_{2}$ was reported in \cite{Park}.   Switching of the ferroelectric polarization in a 1D spin chain via the external magnetic field has been studied experimentally in Ref. \onlinecite{Loidl}. Technically, our  study is straightforwardly applicable  to other non-collinear spin systems, as well.

Due to non-collinearity of spins, the system is entangled.  This entanglement can be exploited  as a resource to enhance the Otto-cycle efficiency.
 The general question concerning a finite output power of the MF quantum Otto engine is still open and will be addressed here.
Using shortcuts to quantum adiabaticity we construct super-adiabatic quantum engine. The paper is organized as follows:  In section II we introduce the system and the theoretical model, in section III we discuss the thermodynamic characteristics of the cycle, and
   in  section IV we will study the coupling of the system to a bath and discuss quantum decoherence phenomena.

\section{Model}
 A one dimensional system with  a charge-driven multiferroicity is modeled well with a chain of $N$ sites (along the $x$ axis) of localized spins
  having frustrated next-nearest neighbor interactions. We apply  a time dependent electric field  $\wp(t)$ which is linearly polarized
  along the $y$ axis, and an external magnetic field $B$  (applied along the $z$  axis). The corresponding Hamiltonian reads
\begin{eqnarray}
\label{Hamiltonian0}
&&\hat{H}_{0}(t)=\hat{H}_{S}+\hat{H}_{SF}(t),\\
&&\hat{H}_{S}=-J_1\displaystyle\sum_{i}\vec{\sigma}_i\cdot\vec{\sigma}_{i+1}-J_2\displaystyle\sum_{i}\vec{\sigma}_i\cdot\vec{\sigma}_{i+2}-\gamma_{e}\hbar B\displaystyle\sum_{i}\sigma_{i}^{z},\nonumber
\end{eqnarray}
 $\hat{H}_{S}$ is  time independent, while $\hat{H}_{SF}$ is  time dependent and contains the coupling of the external electric field to the electric polarization of the chain.
  The exchange coupling  between nearest neighbor spins is chosen ferromagnetic $J_1>0$, while the next-nearest neighbor interaction is antiferromagnetic $J_2<0$.
  The electric polarization  $\vec{P}_{i}$ tagged
  to spin  non-collinearity   reads $\vec{P}_{i}=g_{ME}\vec{e}_{i,\,i+1}\times (\vec{\sigma}_{i}\times\vec{\sigma}_{i+1}) $, where
  $\vec{e}_{i,\,i+1}$ is the unit vector connecting the sites $i$ and $i+1$.  The coupling  strength  of this
 charge-driven magnetoelectric coupling we refer to as $g_{ME}$ (for a detailed discussion of this type of magnetoelectric materials  we refer
 to the reviews \cite{rep_prog_phys} and further references therein).
  The spatially homogeneous, time dependent electric  field  $\wp(t)$ couples to the chain electric polarization $\vec{P}$ such that
  $\vec{\wp}(t)\cdot\vec{P}=d(t)\displaystyle\sum_{i}(\vec{\sigma}_{i}\times\vec{\sigma}_{i+1})^z$, with $d(t)=\wp(t)g_{ME}$. The quantity $(\vec{\sigma}_{i}\times\vec{\sigma}_{i+1})^z$ is known as the $z$ component of the vector chirality.
With this notation   $\hat{H}_{SF}(t)$ reads
\begin{eqnarray}
\label{Hamiltonian01}
&&\hat{H}_{SF}(t)=-\vec{\wp}(t)\cdot\vec{P}=d(t)\displaystyle\sum_{i}(\sigma_{i}^{x}\sigma_{i+1}^{y}-\sigma_{i}^{y}\sigma_{i+1}^{x}).
\end{eqnarray}

    For a first insight we  considered in Eq. (\ref{Hamiltonian0})   four spins, i.e. $N=4$,   which we  solved analytically.
    Thus we can write
    \begin{eqnarray}
     &&\hat{H}_{0}^{N=4}(t)= \sum_{n=1}^{16} \vert\Phi_{n}(\, d(t)\, )\rangle E_n(t) \langle \Phi_{n}(\, d(t)\, )\vert .
     \end{eqnarray}
    The instantaneous state vectors  $\vert\Phi_{n}(d)\rangle$ and energies  $E_n$ are presented in the appendix. Previous studies affirmed weakly pronounced finite-size effects with regard to the efficiency of the cycle \cite{Azimi2} underlining the usefulness of this four-spin working substance.
    In this context we also refer to the remarkable advance in realizing and
    tailoring the chiral magnetic interaction of just few surface deposited atoms by means of spin-polarized scanning tunneling microscopy (cf. Ref. (\onlinecite{wiesendanger}) and references therein). The spin excitation in this case is also captured by the low-energy effective model (\ref{Hamiltonian0}) with appropriately chosen parameters and fields. \\

As mentioned in the introduction, our aim here is to identify adiabaticity shortcuts.
For a  general discussion of  shortcuts to adiabaticity and an overview of the interrelation between the various
 approaches as well as  their historical developments  we refer to the review article (\onlinecite{torron}) and references therein.  Here we will basically  follow
 Berry's transitionless driving formulation\cite{Berry}  which  is equivalent to the counterdiabatic approach
of Demirplak and Rice \cite{Demi}.

%
%
%
%
 %
 %
 %
 %
 %
 %
In the adiabatic approximation a general   state $\vert\Psi_{n}(t)\rangle$ driven by $\hat{H}_{0}(t)$ is cast as
\begin{eqnarray}
\label{states}
&&\vert\Psi_{n}(t)\rangle=\exp\bigg[-\frac{i}{\hbar}\int_{0}^{t}dt^\prime E(t^\prime)\nonumber\\
&&~~~~~~~~~~~~~~~~-\int_{0}^{t}dt^\prime\langle\Phi_{n}(t^\prime )\vert\partial_{t^\prime}\Phi_{n}(t^\prime )\rangle\bigg]\vert\Phi_{n}(t)\rangle.
\end{eqnarray}
With the aid of unitary time-evolution operator
\begin{eqnarray}
\label{Unitary Operator}
&&\hat{U}(t)=\displaystyle\sum_{n}\exp\bigg[-\frac{i}{\hbar}\int_{0}^{t}dt^\prime E(t^\prime)\nonumber\\
&&~~~~~~~-\int_{0}^{t}dt^\prime\langle\Phi_{n}(t^\prime )\vert\partial_{t^\prime}\Phi_{n}(t^\prime )\rangle\bigg]\vert\Phi_{n}(t)\rangle\langle\Phi_{n}(0)|,
\end{eqnarray}
we construct the auxiliary (counter-diabatic)  Hamiltonian
\begin{eqnarray}
\label{auxiliary Hamiltonian}
\hat{H}_{CD}(t)=i\hbar\big(\partial_{t}\hat{U}(t)\big)\hat{U}^\dag(t).
\end{eqnarray}
The reverse state engineering  relies on the requirement that  the states (\ref{states}) solve for the Hamiltonian (\ref{auxiliary Hamiltonian}), meaning that
\begin{eqnarray}
\label{exact solving states}
i\hbar\partial_{t}\vert\Psi_{n}(t)\rangle=\hat{H}_{CD}(t)\vert\Psi_{n}(t)\rangle.
\end{eqnarray}
In this way even for a  fast driving  transitions between eigenstates $\vert\Phi_{n}(t)\rangle$ are prevented.
 After a  relatively simple algebra the counter-diabatic  (CD) Hamiltonian $\hat{H}_{CD}(t)$ takes the form
\begin{eqnarray}
\hat{H}_{CD}(t)=\hat{H}_{0}(t)+\hat{H}_{1}(t),
\end{eqnarray}
where
\begin{eqnarray}
\label{Hamiltonian1}
\hat{H}_{1}(t)=i\hbar\displaystyle\sum_{m\neq n}\frac{\vert\Phi_{m}\rangle\langle\Phi_{m}\vert\partial_{t}\hat{H}_{0}(t)\vert\Phi_{n}\rangle\langle\Phi_{n}\vert}{E_n - E_m}.
\end{eqnarray}
We adopt the initial conditions for the driving protocol as $\hat{H}_{CD}(0)=\hat{H}_{0}(0)$, $\hat{H}_{CD}(\tau)=\hat{H}_{0}(\tau)$. Thus,
 on the time interval $t\in [0, \tau]$ we achieve a fast adiabatic dynamics by means of the counter-diabatic Hamiltonian $\hat{H}_{CD}(t)$. Taking into account (\ref{Hamiltonian0})-(\ref{Hamiltonian1}) after laborious but straightforward calculations we obtain
\begin{eqnarray}
\label{Hamiltonian1final}
\hat{H}_{1}(t)=i\hbar A(t)\big(\vert\Phi_{6}(t)\rangle\langle\Phi_{7}(t)\vert-\vert\Phi_{7}(t)\rangle\langle\Phi_{6}(t)\vert\big).
\end{eqnarray}
The  explicit form of the time dependent parameter we derived as  $A(t)=\frac{4\dot{d}(t)(\lambda+\mu)\alpha\nu}{d(t)(\lambda-\mu)}$ (please see the appendix for
the  determining equations of $\alpha, \lambda, \mu,  \nu$). Note that the time dependence in the model appears through the external electric field $d(t)$ and its time derivative $\dot{d}(t)$. The obtained explicit form of the functions (\ref{states}) and counter-diabatic Hamiltonian $\hat{H}_{CD}(t)$ is rather involved and is presented in the appendix.

\section{Thermodynamic characteristics of the cycle}

In recent years there have been a  growing interest in the non-equilibrium statistical physics, especially for constrained and finite-size systems, as these are becoming feasible and controllable experimentally \cite{Scully,Georgescu,Rossnagel}.
An important point in this context is that, while fluctuations are ignorably small in macroscopic systems they become important for small systems, particularly  in  a non-equilibrium situation. The physics in this case cannot be captured
by conventional equilibrium statistical mechanics and equilibrium thermodynamics. Pioneering works in this direction were done for instant by
 G. N. Bochkov and Yu. E. Kuzovlev  \cite{Bochkov} (see the review paper \cite{Kuzovlev} and references therein) and are receiving a renewed interest w
  with the rise of nanotechnology rendering possible the realization and test of theoretical concept.
  Quantum heat engines for example were  proposed (and realized) as portable  nano ``thermodynamic'' circuits to produce  useful work on the nanoscale.  Finite fluctuations being  inherent to quantum heat engines  should therefore be carefully addressed \cite{Jarzynski2}.

\begin{figure}[h]
\includegraphics[width=0.4\textwidth]{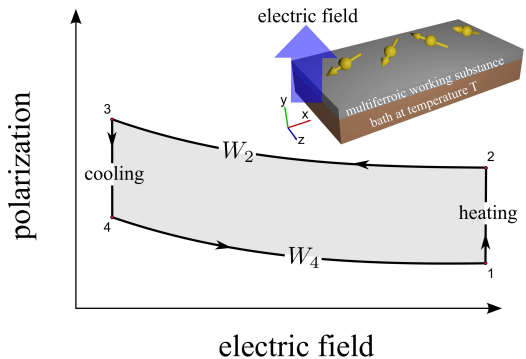}
\caption{\label{sketch}
The  cycle under study  with a chiral multiferroic working substance. It has four strokes.The  isochoric processes are
 from $1\rightarrow 2$ and from $3\rightarrow 4$.
 The processes $2\rightarrow 3$ and $4\rightarrow 1$ are  quantum adiabatic. We vary the amplitude of the electric field from $\Delta E_{n}=E_{n}(d_{0})-E_{n}(d_{1})$
and  the working substance performs during
  $2\rightarrow 3$ a positive magnetoelectric work. }
\end{figure}

The quantum Otto cycle consists of two quantum isochoric and two adiabatic strokes. The quantum isochoric strokes correspond to heat exchange between the working substance and the cold and the hot heat baths. During the quantum isochoric strokes the level populations are altered. The  MF working substance produces work during the adiabatic process.  Changing  the amplitude of the applied external electric field modifies the energy spectrum of the system. This is the mechanism behind producing work. The quantum Otto cycle and the  MF-based engine are   detailed in recent work \cite{Azimi2}. Here we concentrate on the thermodynamic characteristics such as: the output power of the cycle and the irreversible work. We choose a particular type of the time dependence for the external electric field
\begin{eqnarray}
\label{protocol}
d(t)=\epsilon\bigg(\frac{t^3}{3\tau}-\frac{t^2}{2}\bigg)+d_0.
\end{eqnarray}
 The working parameter (i.e., the electric field) $d(t)$ during the adiabatic strokes varies from $d_0\longrightarrow d_1$ (stroke $2\rightarrow 3$) and $d_1\longrightarrow d_0$ (stroke $4\rightarrow 1$). The scheme of the cycle is sketched in  Fig.~\ref{sketch}. From Eq. (\ref{exact solving states})-(\ref{Hamiltonian1final}) it is evident that in this case the requirement for the shortcuts of adiabaticity  $$\hat{H}_{CD}(0)=\hat{H}_{0}(0),\: \hat{H}_{CD}(\tau)=\hat{H}_{0}(\tau),$$ is fulfilled. The chosen driving protocol Eq.(\ref{protocol})
should satisfy some restrictive constraints imposed by the adiabatic shortcuts. On the other hand, the protocol should be experimentally accessible and amenable to theoretical interpretations. For a finite time thermodynamic process, the output power of the quantum Otto cycle can be written as\cite{Altintas,Esposito2010,delCampo2014}
\begin{eqnarray}
\Re=\frac{-\big(\langle W_2\rangle_\mathrm{ad}+\langle W_4\rangle_\mathrm{ad}\big)}{\tau_1(T_H)+\tau_2+\tau_3(T_L)+\tau_4}.
\end{eqnarray}
Here $\tau_1(T_H)$, $\tau_3(T_L)$ are the relaxation times of the MF working substance in contact with the hot and the cold thermal baths (strokes $1\rightarrow 2$ and $3\rightarrow 4$), $\tau_2$ and $\tau_4$ correspond to the duration of the adiabatic strokes, $\langle W_2\rangle_\mathrm{ad}$ and $\langle W_4\rangle_\mathrm{ad}$ correspond to the work produced during the quantum adiabatic strokes. The condition
\begin{eqnarray}
&&\langle W_2\rangle_\mathrm{ad}+\langle W_4\rangle_\mathrm{ad}+Q_\mathrm{in}+Q_\mathrm{out}=0,\nonumber
\end{eqnarray}
during the whole cycle should be satisfied. The corresponding absorbed heat $Q_\mathrm{in}$ and the released heat $Q_\mathrm{out}$ by the working substance are defined as follows\cite{Altintas}
\begin{eqnarray}
Q_\mathrm{in}&=& \sum_{n}E_{n}(0)\bigg(\frac{e^{-\beta_{H}E_n(0)}}{\sum_{n}e^{-\beta_H E_n(0)}}-\frac{e^{-\beta_{L}E_n(\tau)}}{\sum_{n}e^{-\beta_L E_n(\tau)}}\bigg),\nonumber\\
Q_\mathrm{out}&=&\sum_{n}E_{n}(\tau)\bigg(\frac{e^{-\beta_{L}E_n(\tau)}}{\sum_{n}e^{-\beta_L E_n(\tau)}}-\frac{e^{-\beta_{H}E_n(0)}}{\sum_{n}e^{-\beta_H E_n(0)}}\bigg).\nonumber\\
\label{heats}\end{eqnarray}
Irreversibility of classical thermodynamical processes are quantified in terms of Clausius inequality
\begin{eqnarray}
&&\Delta S=S_\mathrm{re}+S_\mathrm{ir},
\end{eqnarray}
where
 $S_\mathrm{re}=\beta Q$ is the equilibrium entropy, $Q$ is the transferred heat, and $\beta=1/T$ is the inverse temperature.  For irreversible processes $S_\mathrm{ir}>0$. In  quantum thermodynamics the situation is more delicate. E.g., the concept of  work for mesoscopic systems has been revisited recently \cite{Jarzynski1,delCampo2014}. The work performed on a finite quantum system  is not an observable but a randomly distributed quantity \cite{Jarzynski2, Deffner}. Any sudden abrupt change, fast driving or a quench drags  the system into a non-equilibrium state. Hence, recipes of the equilibrium thermodynamics need to be questioned.
  A fast transformation leads to the ``parasitic'' irreversible work $\Delta S_\mathrm{ir}=\beta \big<W_\mathrm{ir}\big>$ which amounts to the  difference between the total work and the change of the free energy $ \big<W_\mathrm{ir}\big>= \big<W\big>-\Delta F$. The expression for the total quantum mean work has been deduced in \cite{Deffner} and reads
  \begin{eqnarray}
  &&\big<W\big>=\sum_{n,m}\big(E_{n}(t)-E_{m}(0)\big)P_{mn}(t)P_{m}^{(0)}(\beta),
  \end{eqnarray}
   where  $$P_{mn}(t)=\vert\langle\Phi_{n}(t)\vert\hat{U}(t)\vert\Phi_{m}(0)\rangle\vert^2,$$ is the transition probability between the eigenstates of the Hamiltonian $\hat{H}_{0}(t)$, and $P_{m}^{(0)}(\beta)$ describes the level populations in equilibrium at the temperature $\beta$.
If the cycle is reversible (by virtue of a realized counter-diabatic driving) at the end of the stroke the transition probability simplifies to $P_{mn}(\tau)=\delta_{mn}$. Therefore, the  expressions of the adiabatic work for the cycle strokes are

\begin{eqnarray}
\label{Works}
&&\langle W_2\rangle_\mathrm{ad}=\displaystyle\sum_{n}\big[E_n(\tau)-E_n(0)\big]P_{n}^{(1)}(\beta_H),\nonumber\\
&&\langle W_4\rangle_\mathrm{ad}=\displaystyle\sum_{n}\big[E_n(0)-E_n(\tau)\big]P_{n}^{(3)}(\beta_L).
\end{eqnarray}
 $P_{n}^{(1)}(\beta_H)=\frac{e^{-\beta_{H}E_n(0)}}{\sum_{n}e^{-\beta_H E_n(0)}}$, $P_{n}^{(3)}(\beta_L)=\frac{e^{-\beta_{L}E_n(\tau)}}{\sum_{n}e^{-\beta_L E_n(\tau)}}$ are the level populations in equilibrium at the temperatures $\beta_H=1/T_H$, $\beta_L=1/T_L$ respectively.
Thus, the criteria for the quantum adiabaticity, i.e.  the success of the counter-diabatic driving, is  the vanishing of the irreversible work at the end of the adiabatic stroke $\beta \big<W_\mathrm{ir}\big>=0$. Therefore, along  with the total mean work and the mean square fluctuations of the total work we will study the irreversible work as well.
The explicit form of (\ref{Works}) after taking into account (\ref{Unitary Operator})-(\ref{Works}) is presented in the appendix $(A6)$,$(A7)$. For the partition functions we introduced the following notations
\begin{eqnarray}
&&Z=\sum_{n}e^{-\beta_H E_n(0)}, \mbox{ and } Z^{\prime}=\sum_{n}e^{-\beta_L E_n(\tau)}.
\end{eqnarray}
To quantify the mean square fluctuations
 \begin{eqnarray}
 \label{fluctuations1}
 \Delta W_\mathrm{ad}={\big[\langle W^2\rangle_\mathrm{ad}-\langle W\rangle_\mathrm{ad}^2\big]}^{\frac{1}{2}},
 \end{eqnarray}
 for the work $W=W_2 + W_4$  we utilize the following ansatz
\begin{eqnarray}
&&\langle W^2\rangle_\mathrm{ad}=\langle W_2^2\rangle_\mathrm{ad}+\langle W_4^2\rangle_\mathrm{ad}+2\langle W_2\rangle_\mathrm{ad}\langle W_4\rangle_\mathrm{ad},\nonumber\\
&&\langle W\rangle_\mathrm{ad}^2=\langle W_2\rangle_\mathrm{ad}^2+\langle W_4\rangle_\mathrm{ad}^2+2\langle W_2\rangle_\mathrm{ad}\langle W_4\rangle_\mathrm{ad}.
\end{eqnarray}
The mean values of the work are defined as
\begin{eqnarray}
\label{Works2}
\langle W_{2,(4)}^2\rangle_\mathrm{ad}&=&
\displaystyle\sum_{n}{\big[E_n(\tau)-E_n(0)\big]}^2P_{n}^{(1),(3)}(\beta_H,\beta_L),\nonumber\\
\langle W_{2,(4)}\rangle_\mathrm{ad}^2&=&
{\bigg(\displaystyle\sum_{n}\big[E_n(0)-E_n(\tau)\big]P_{n}^{(1),(3)}(\beta_H,\beta_L)\bigg)}^2.\nonumber\\
\end{eqnarray}
With these relations, the means of the square components $\langle W_{2,(4)}^2\rangle_\mathrm{ad}$ are presented in the appendix $(A8)$, $(A9)$.
 For the square of the mean values $\langle W_{2,(4)}^2\rangle_\mathrm{ad}$ we employ the square of Eq. $(A6)$, $(A7)$ in the appendix.

As mentioned, for  finite systems the concept of  work need to be revisited. The fluctuations come into play and as a result the non-equilibrium work is different from the equilibrium work\cite{Jarzynski1,delCampo2014}. For a  further discussion we will introduce the quantum Kullback-Leibler divergence $$S\big(\varrho_{A}\parallel \varrho_{B}\big)= Tr \big(\varrho_{A}
\ln\varrho_{A}-\varrho_{A}\ln\varrho_{B}\big),$$ and rewrite the expression for the irreversible work in the following form \cite{Deffner}
\begin{eqnarray}
\label{non-equilibrium work}
\big<W_\mathrm{ir}\big>= \big<W\big>-\Delta F=\frac{1}{\beta}S\big(\rho_t\|\rho_t^{\mathrm{eq}}\big).
\end{eqnarray}
Here we introduced the following notations
\begin{eqnarray}
&&S\big(\rho_t\|\rho_t^{\mathrm{eq}}\big)=-\displaystyle\sum_{n,k}P_n^0 P_{kn}^t \ln P_k^t+\displaystyle\sum_{n}P_n^0\ln P_n^0,\nonumber\\
&&\Delta F=-\frac{1}{\beta}\ln\bigg(\frac{\sum_{n}\exp\big[-\beta E_{n}(t)\big]}{\sum_{m}\exp\big[-\beta E_{m}(0)\big]}\bigg).
\end{eqnarray}
where $$P_n^0=\exp[-\beta E_n]/\sum_{n}\exp[-\beta E_n],$$ and $$P_k^t=\exp[-\beta E_k^{CD}]/\sum_{k}\exp[-\beta E_k^{CD}],$$ correspond to the level populations and $P_{kn}^t=|\langle\Psi_{n}(t)|\hat{U}(t)|\Phi_{k}(0)\rangle|^2$ to the transition amplitudes.
For an insight into the analytical results we present plots of the thermodynamic quantities. We adopt dimensionless parameters $$J_1=1,\, J_2=-1,\, B=0.1,\, d_0=2.5,\, \epsilon=1.$$ In real units these parameters correspond to the one phase MF material\cite{Park}  $\textrm{LiCu}_{2}\textrm{O}_{2}$,   $J_1=-J_2=44[\mathrm{K}]$. The external driving fields strengths are $B=3[\mathrm{T}]$, $\wp=5\times10^{3}[\mathrm{kV/cm}]$.
We assume that the duration of the adiabatic strokes of the cycle are equal to $\tau_2=\tau_4=\tau$. The time unit in our calculations corresponds to the $\hbar/J_{1}\approx 0.1[\mathrm{ps}]$. CD driving allows reducing  the driving time. Implementing a short driving protocol is supposed  to maximize the output power of the cycle.
In order to calculate the thermal relaxation times $\tau_1(T_{H}),~\tau_3(T_{L})$ in the next section we  solve self-consistently the
Lindblad master equation. Our calculations (see bellow) show that the relaxation times are shorter than the duration of the implemented adiabatic strokes $\tau_1(T_{H}),~\tau_3(T_{L})\ll\tau$. Therefore, in the first approximation we neglect the relaxation times when calculating the output power.


\begin{figure}[h]
\includegraphics[width=0.4\textwidth]{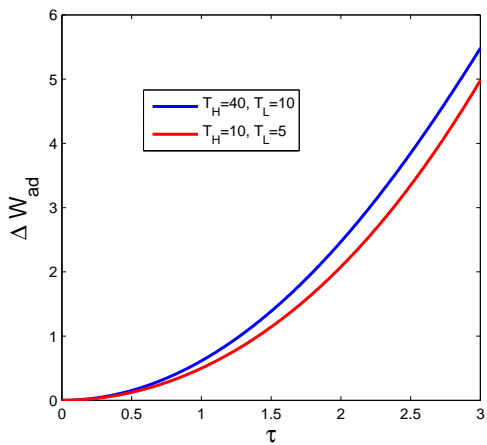}
\caption{\label{rms} Standard deviation of the work $\Delta W_\mathrm{ad}$ in scaled units for two different heat and cold bath temperatures.
 The other parameters are: $\varepsilon =1$, $J_1=1, J_2=-1, B=0.1, d_0=2.5$. Unscaled unit of $\Delta W_{ad}$ amounts  to $6\times10^{-22}[J]$.}
\end{figure}

\begin{figure}[h]
  \includegraphics[width=0.40\textwidth]{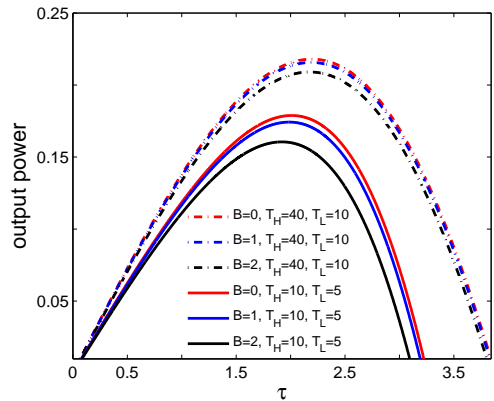}
 \caption{\label{power} Output power for different values of the magnetic filed $B$, heat and cold bath temperatures. The other parameters in scaled units are $J_1=1, J_2=-1, d_0=2.5$, $\varepsilon =1$. In unscaled
 units the parameters correspond to $\wp_0=5\times10^{3}[\mathrm{kV/cm}]$ and time unit is $0.1[ps]$. Unit of the power is  $6\times10^{-9}[W]$. }
 \end{figure}

\begin{figure}[h]
\includegraphics[width=0.4\textwidth]{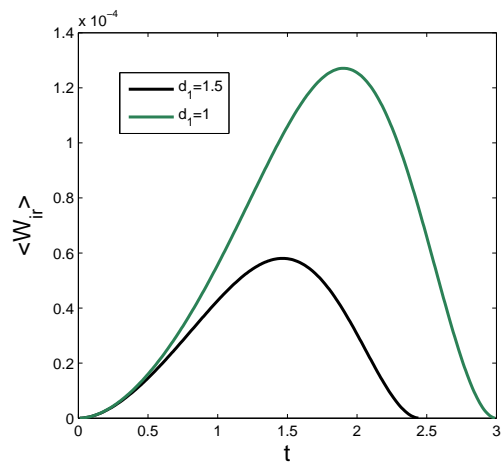}
\caption{\label{irWork} $\big<W_\mathrm{ir}\big>$ for the values of parameters $J_1=1, J_2=-1, B=0.1, d_0=2.5$.
Unscaled unit of  $\big<W_\mathrm{ir}\big>$ is  $6\times10^{-22}[J]$. This figure quantifies the irreversible work accumulated  during the performance of the adiabatic stroke. Because of the implemented adiabatic shortcut, at the end of the stroke the irreversible work vanishes.}
\end{figure}

As evident from Fig.~\ref{rms} the work mean square fluctuations increase with the stroke duration $\tau$. We also infer  that the fluctuations increase with temperature. The modulation depth of the driving parameter $d(t,\tau)=\epsilon \big(t^{3}/3\tau -t^{2}/2\big)+d_0,~\dot{d}(0,\tau)=\dot{d}(\tau,\tau)$ enhance the work mean square fluctuations for longer duration of the adiabatic strokes $\tau$. The cycle duration enhances as well however this has an adverse effect on the output power (See Fig.~\ref{power}). These two factors compete resulting in the optimal time length of the adiabatic strokes $\tau_{op} =0.23 [\mathrm{ps}]$.
For the irreversible work (Fig.~\ref{irWork}) we again have a non-monotonic behavior. For larger times the system tends to equilibrium.

We observed (cf. Fig.~\ref{power}) that a strong magnetic field is counterproductive for the output power.

\section{Efficiency of the engine and finite-size effects}

Naturally the work produced by the engine and the output power increase with the size of the working substance. In contrast, the  situation regarding  the efficiency might be counterintuitive. In this section we present results about the dependence of the cycle efficiency on the length of the MF chain.

For the efficiency of the engine we use standard expression:
\begin{eqnarray}
	&&\eta = \frac{\delta W}{\delta Q_\mathrm{in}}.
\end{eqnarray}
Here $\delta W$ corresponds to the work produced by engine and $\delta Q_\mathrm{in}$
quantifies heat transferred from the hot bath to the working substance.

\begin{figure}[h]
 \includegraphics[width=0.95\columnwidth]{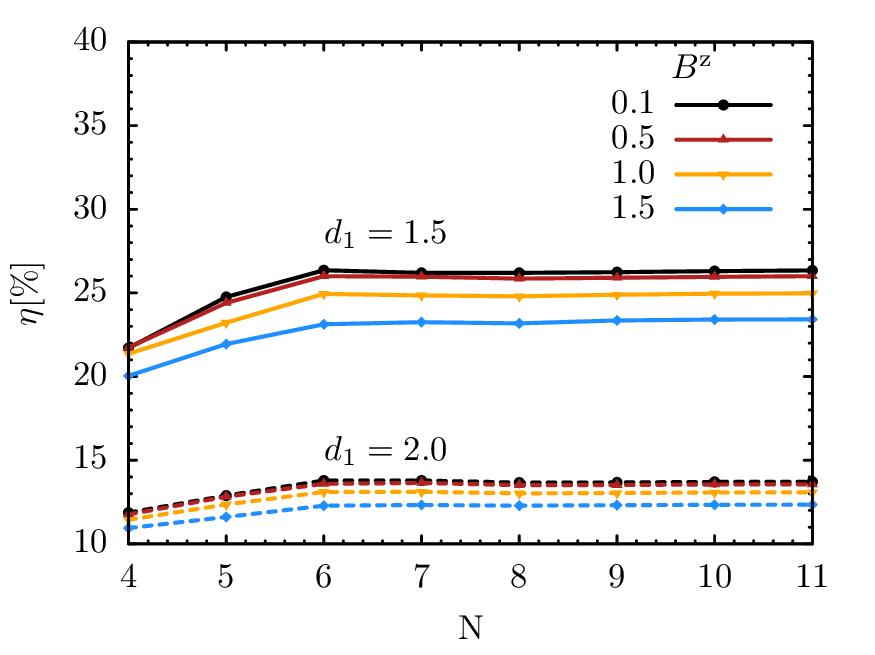}
  \caption{\label{fig:effi} Dependence of the cycle efficiency on the length of the working substance. For different values of the electric and magnetic fields. As evident,   increasing the chain length the efficiency of the cycle undergoes a rapid transition to the saturated value
and remains constant. The following set of parameters were used $J_1 = -1$, $J_2 = 1$, $d_0 = 2.5$. The solid lines correspond to $d_1 = 1.5$. The dashed lines correspond to $d_1 = 2.0$ and $B$ as in the figure.}
\end{figure}

At first we examine finite system gradually increasing number of the spins. In case of a finite system expressions for produced work $<W_{2}>_{ad}$ and transferred heat $Q_{in}$ are defined in Eq. (13), (16). Our numerical results show that (see Fig. \ref{fig:effi})
with increasing the chain length the efficiency of the cycle undergoes a rapid transition to a saturated value and stays constant. Therefore, on the mesoscopic scale we do not expect prominent changes in the cycle efficiency.

In the thermodynamic limit, when the length of the chain tends to  infinity,  the energy spectrum of the system becomes continuous. We assume that the  chirality term $d \sum\limits_n \left( \hat{\sigma}_n \times \hat{\sigma}_{n+1} \right)_z$ is much weaker as compared to the exchange interaction $d \ll J_{1},J_{2}$. This assumption is valid
if the electric field is not too strong. Then spectral properties of the quasi-particle excitations in the system are quantified via the following dispersion relation \cite{ChenBo}:
\begin{equation}
\label{dispersion relation}
	\omega_q \left(d\right) = \sqrt{A^2 \left(q\right) - B^2 \left( q \right)} + 4d \sin \left( q \right) \, .
\end{equation}
Here $\cos Q = -J_1 / 4 J_2$ and we introduced the following notations $A\left( q \right) = J_1 \left(-2 \cos Q  + \left(1 +\cos Q \right) \cos q  \right)
+ J_2 \left(-2 \cos 2Q + \left(1 +\cos 2Q\right) \cos 2q \right)$ and $B\left( q \right) = J_1 \left(\cos Q - 1 \right) \cos q
					 + J_2 \left(\cos Q - 1 \right) \cos 2q$. The last term in Eq.(\ref{dispersion relation}) corresponds to the contribution of the magnetoelectric coupling.

The free energy of the MF working medium in the thermodynamic limit reads
\begin{equation}
	F \left( d \right) = T \sum\limits_q \ln \left( 1 - \exp\left(-\frac{\hbar \omega_q}{T k_B}\right)\right)\, .
\end{equation}
The work produced by engine is equal to the change of free energy.
\begin{equation}
	\delta W = \Delta F = F \left( d_1 \right) - F \left(d \right).
\end{equation}
After a little algebra for the total internal energy of the working substance $U = -T^2 \frac{\partial}{\partial T} \left(\frac{F}{T}\right)$
we deduce
\begin{equation}
	U
	= \int\limits_{0}^{\pi} \frac{\hbar \omega_q \exp\left(-\hbar \omega_q / T k_B\right)}{1-\exp\left(-\hbar \omega_q / T k_B\right)} dq .
\end{equation}
The heat transferred to the engine in the thermodynamic limit is defined via:
\begin{eqnarray}
	\delta Q_\mathrm{in} = \delta U = \int\limits_{0}^{\pi} && \frac{\hbar \omega_q \exp\left(-\hbar \omega_q / T_H k_B\right)}{1-\exp\left(-\hbar \omega_q / T_H
 k_B\right)} \nonumber\\
-&&\frac{\hbar \omega_q \exp\left(-\hbar \omega_q / T_L k_B\right)}{1-\exp\left(-\hbar \omega_q / T_L k_B\right)}dq .
\end{eqnarray}

After substituting Eq. (26) and E.(28) in the Eq. (23) we plot efficiency of the engine in the thermodynamic limit.

\begin{figure}[h]
 \includegraphics[width=0.95\columnwidth]{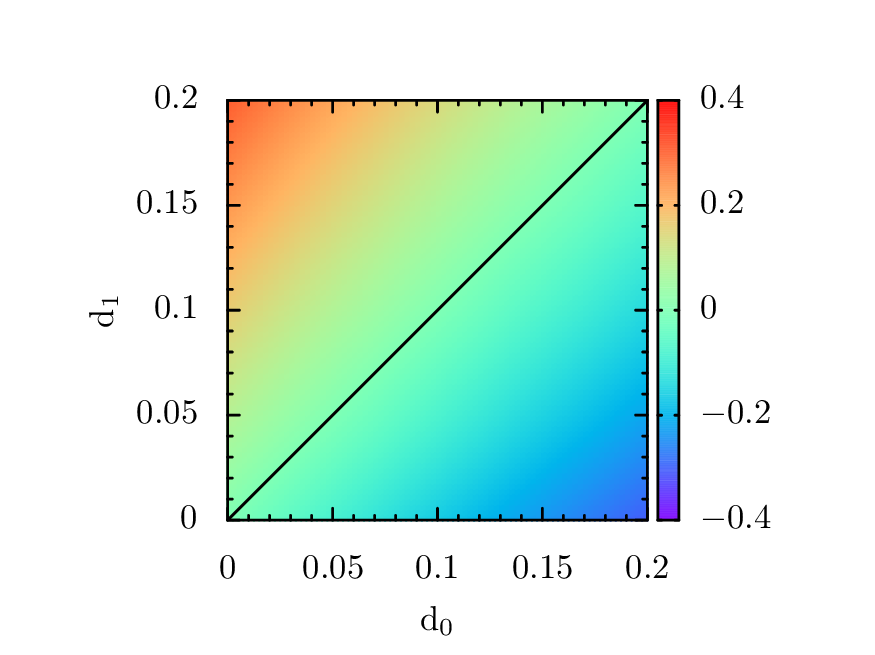}
  \caption{\label{fig:effi_inf}  Efficiency of the engine in the thermodynamic limit. Set of parameters $J_1 = -1$, $J_2 = 1$, $B = 0.1$. The black solid line is the border which separates two domains of parameters ($d_{0},d_{1}$) for which engine works as a heat engine or refrigerator.}
\end{figure}

As we see from the Fig. \ref{fig:effi_inf} the efficiency of the engine can be negative as well. This means that for this particular choice of the parameters the engine is working as a refrigerator. One can switch from the heat engine to the refrigerator regime by replacing the parameters $d_{0} \rightarrow d_{1}$.

\begin{figure}[h]
\includegraphics[width=0.4\textwidth]{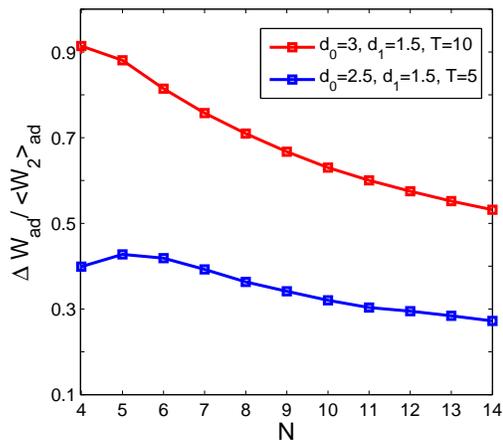}
\caption{\label{DeltaW_size} $\Delta W_\mathrm{ad}/\langle W_{2}\rangle_\mathrm{ad}$ as a function of the system size $N$. The other parameters are $J_1=1, J_2=-1, B=0.1$.}
\end{figure}

\begin{figure}[h]
\includegraphics[width=0.4\textwidth]{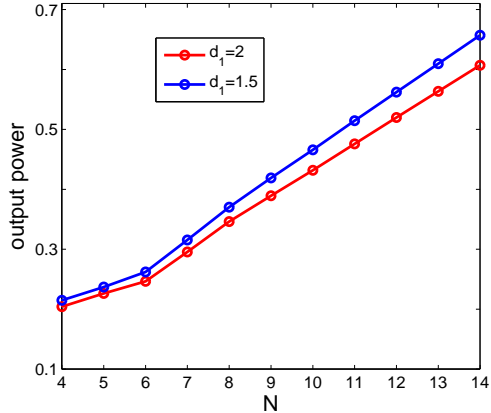}
\caption{\label{Power_size} Output power as a function of the system size $N$ for $J_1=1, J_2=-1, B=0.1, d_0=2.5, T_H=40, T_L=10$.}
\end{figure}

Not only efficiency but other thermodynamic characteristics of the system may show nontrivial interesting finite size effects. The first quantity of our interest is the ratio between mean square fluctuations of the produced work $\Delta W_\mathrm{ad}={\big[\langle W^2_{2}\rangle_\mathrm{ad}-\langle W_{2}\rangle_\mathrm{ad}^2\big]}^{\frac{1}{2}}$ and produced work itself $\Delta W_\mathrm{ad}/\langle W_{2}\rangle_\mathrm{ad}$. As we see from the Fig. \ref{DeltaW_size} this ratio gradually decreases with a system's size. Also we see that fluctuations become smaller at lower temperature.
Output power of the engine as expected is increasing with the size of the working substance see Fig. \ref{Power_size}. This result is clear because
produced work increases with the number of spins contributing in the work.

\section{Entanglement and efficiency of the cycle}

\begin{figure}[h]
\includegraphics[width=0.45\textwidth]{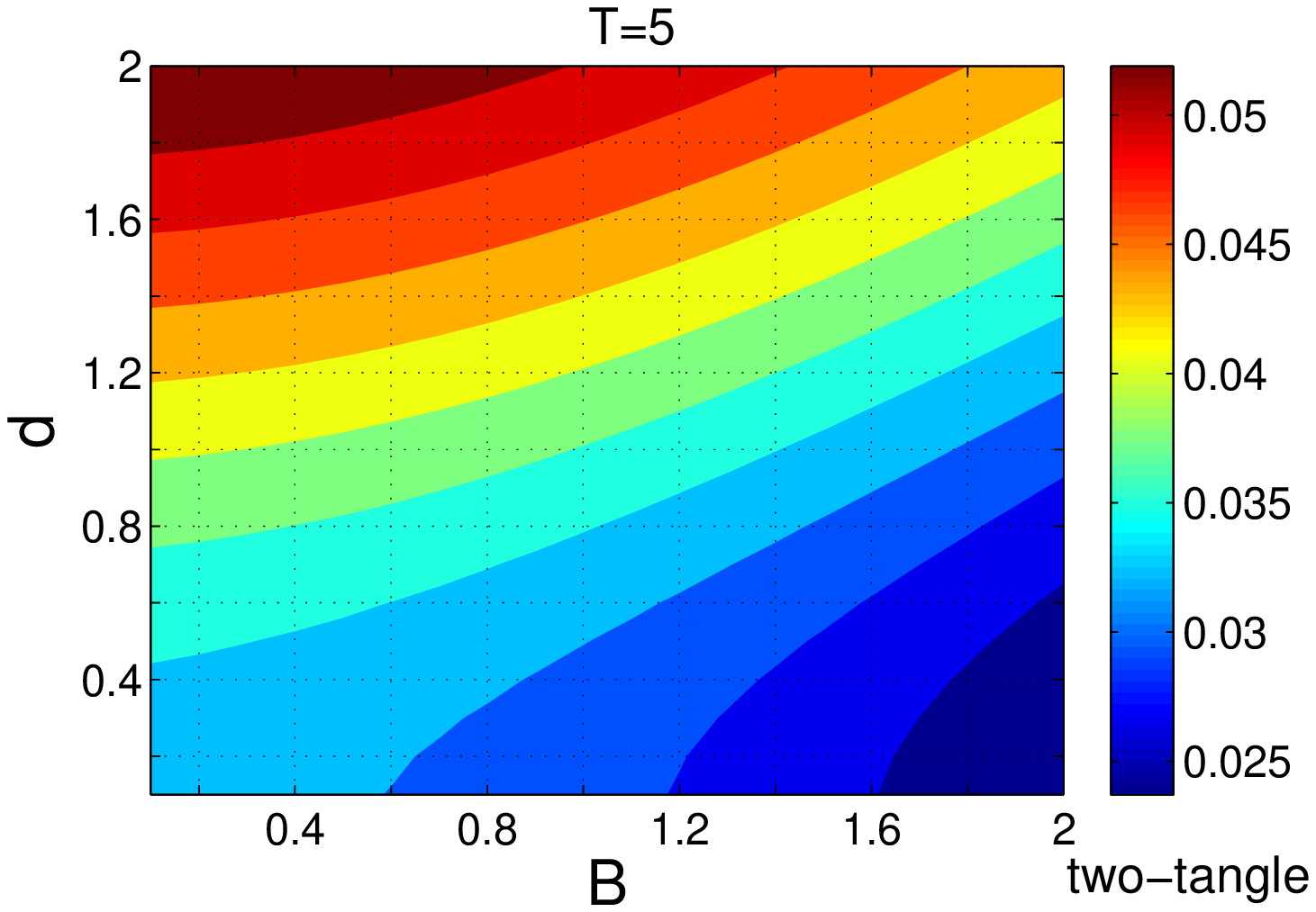}\
\includegraphics[width=0.45\textwidth]{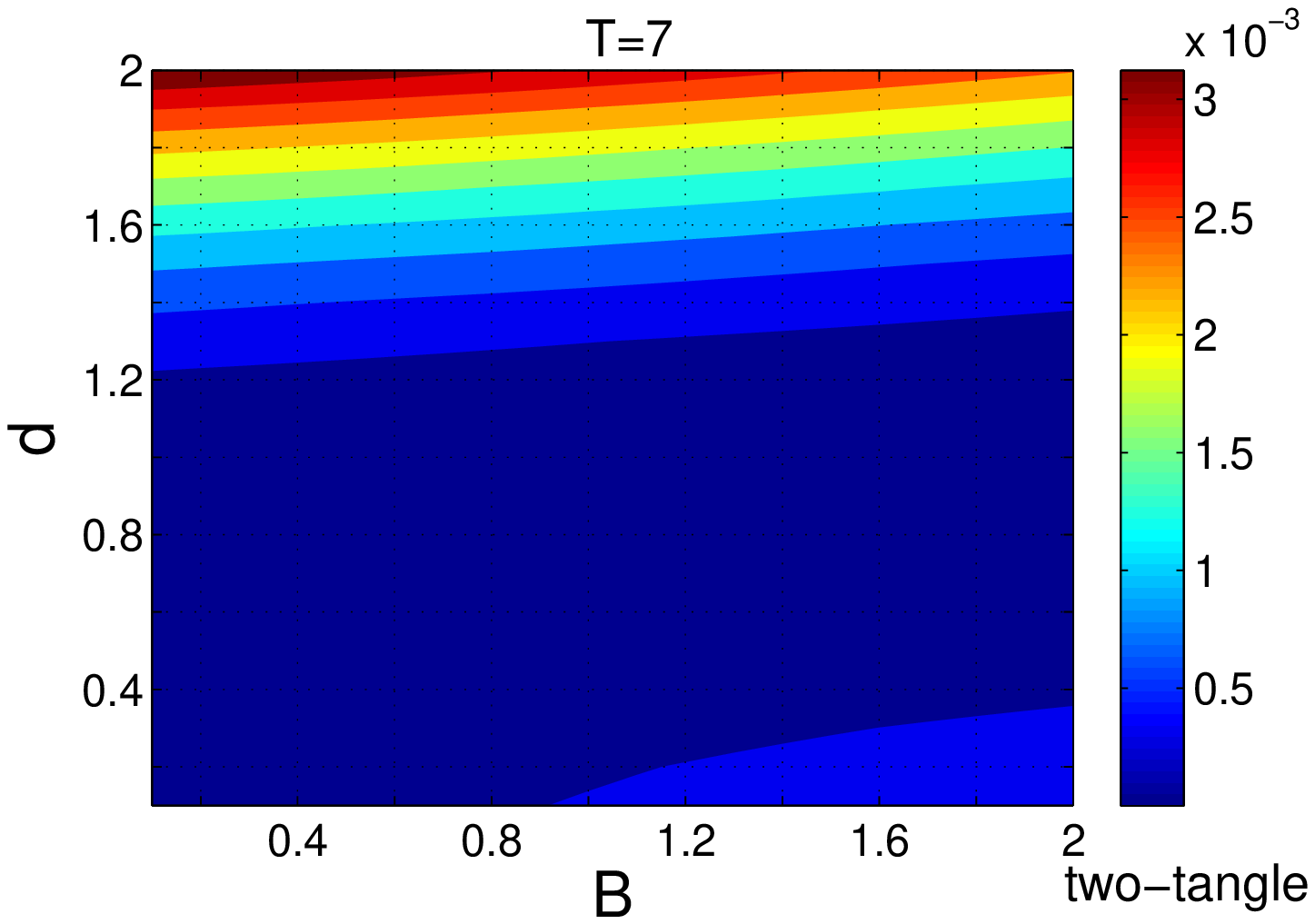}
\caption{\label{contourplots1} Two-tangle entanglement
 as a function of electric and magnetic fields $d$ and $B$.}
\end{figure}

\begin{figure}[h]
 \includegraphics[width=0.45\textwidth]{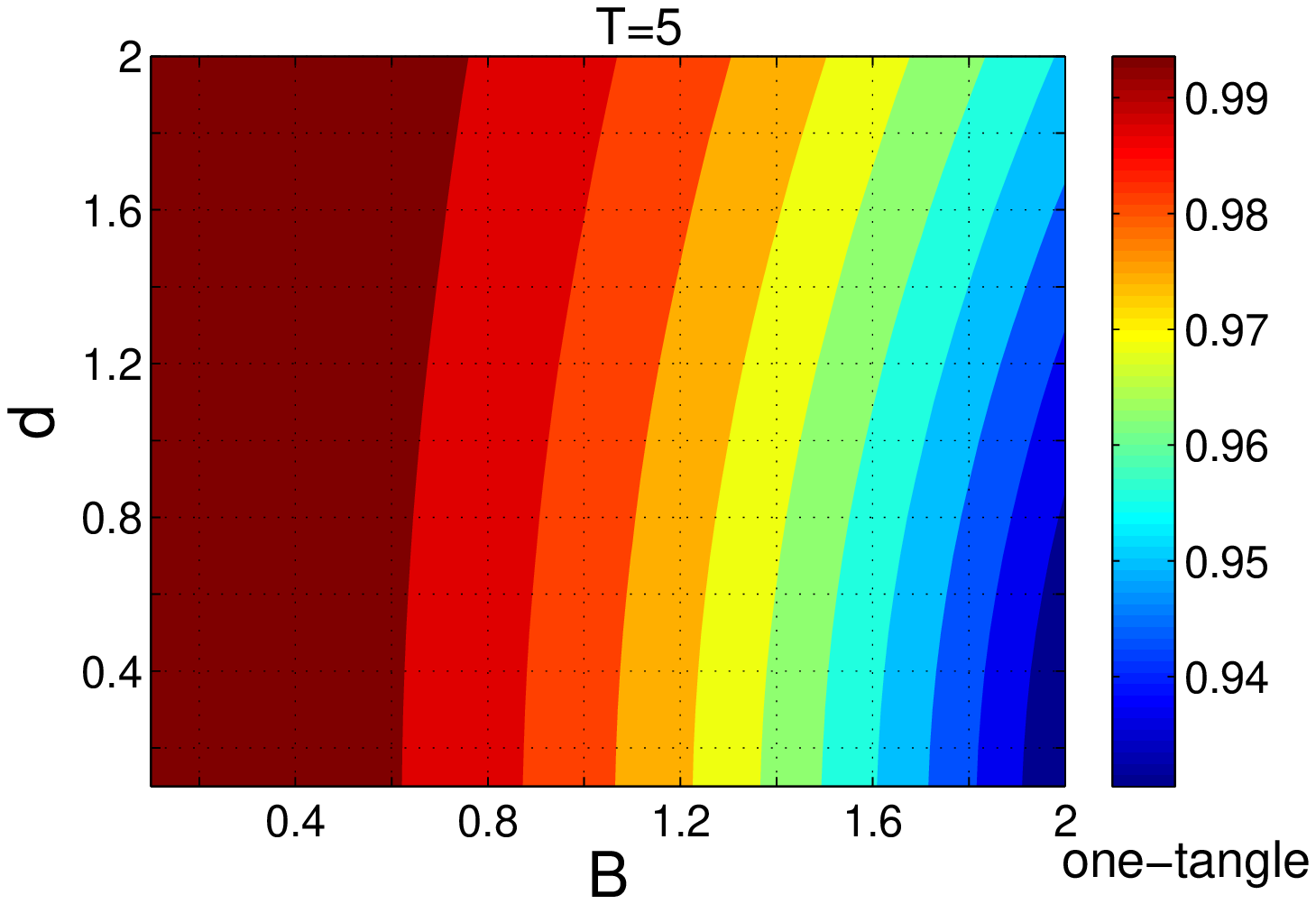} \
 \includegraphics[width=0.45\textwidth]{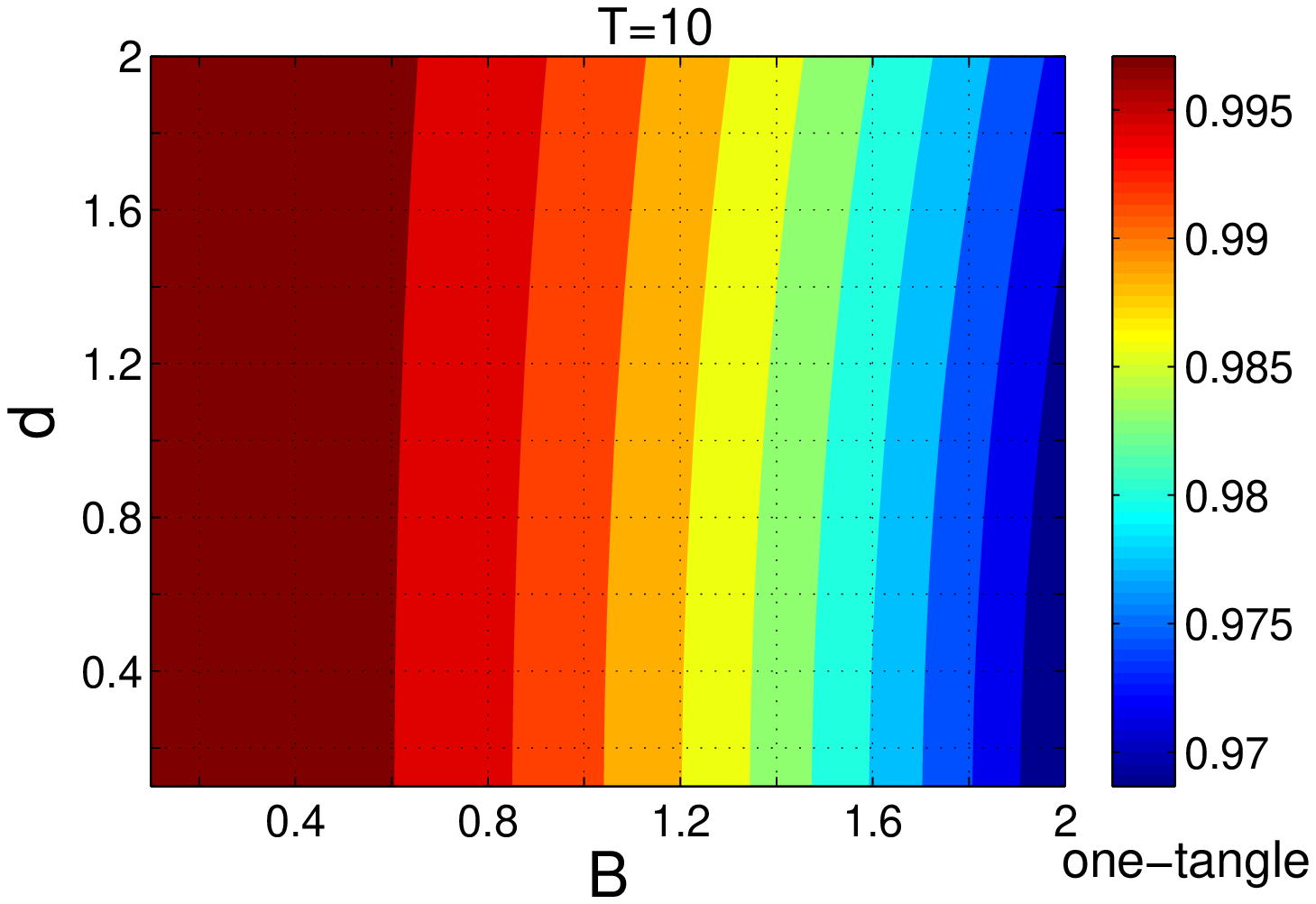}
 \caption{\label{contourplots2} One-tangle entanglement
 as a function of electric and magnetic fields $d$ and $B$.}
\end{figure}

For strong B-field the system is driven from the (entangled) chiral to the (product) collinear state. Hence, it is of relevance to inspect the connection  between the entanglement and the produced work. Since we are interested in the thermal entanglement we will consider states thermalized with hot and cold baths.
Quantum entanglement can be local and nonlocal, shared by two particles only, or  by the whole system.
The pair entanglement is quantified in terms of the two tangle $\tau_{2}$.  The one tangle $\tau_{1}$ measures the many-body entanglement.
These quantities are defined as  \cite{Amico}
\begin{eqnarray}
&&\tau_2=\sum_{m}^{4}C_{nm}^{2},\nonumber\\
&&C_{nm}=\mathrm{max}(0,\sqrt{R_{nm}^{(1)}}-\sqrt{R_{nm}^{(2)}}-\sqrt{R_{nm}^{(3)}}-\sqrt{R_{nm}^{(4)}}),\nonumber\\
&&\tau_1=4{\rm det }\rho_1.
\end{eqnarray}
 $C_{nm}$ is the pair concurrence between the spins on the sites  $n$ and $m$  and $R_{nm}$ are the eigenvalues of the matrix
$R_{nm}=\rho_{nm}^{R}(\sigma_{1}^{y}\bigotimes\sigma_{2}^{y})(\rho_{nm}^{R})^{*}(\sigma_{1}^{y}\bigotimes\sigma_{2}^{y})$.
$\rho_{nm}^{R}$ and $\rho_1$ are respectively the two spins and the single spin reduced density matrices,  which are obtained from the density matrix of the total system $\hat{\rho}$. As the output power, the pair entanglement  is also larger for  weaker magnetic fields  Fig.~\ref{contourplots1} which underlines the interrelation between the pair entanglement and the output power.
The many-body entanglement shows a more robust behavior Fig.~\ref{contourplots2} with increasing the magnetic field.

\begin{figure}[h]
\includegraphics[width=0.4\textwidth]{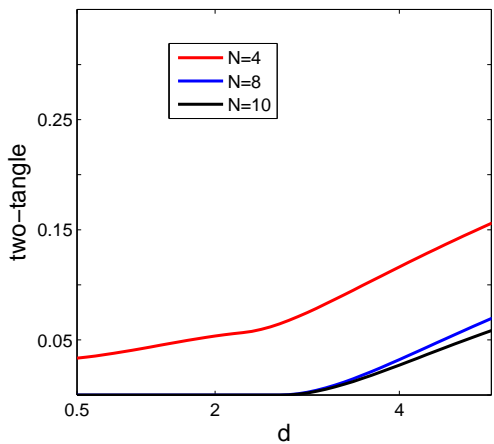}
\caption{\label{Twotangle_size} Two-tangle as a function of the electric field $d$ for three different size $N$ of the system. The parameters are $J_1=1, J_2=-1, B=0.1, d_0=2.5, T=5$.}
\end{figure}

\begin{figure}[h]
\includegraphics[width=0.4\textwidth]{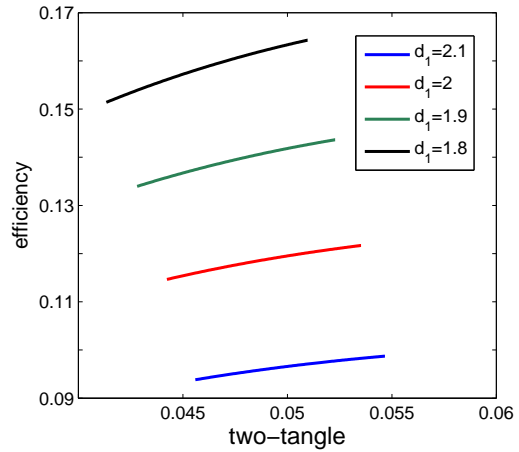}
\caption{\label{Effi-Twotangle_4} The efficiency of the cycle as a function of the two-tangle. Magnetic field is varied between $0.1<B<2$. The other parameters are $J_1=1, J_2=-1, B=0.1, d_0=2.5, T_H=10, T_L=5$.}
\end{figure}

We observed that pair entanglement and local correlations (two tangle $\tau_{2}$) are stronger in small system and drastically decays with the system's size Fig. \ref{Twotangle_size}. We also clearly see connection between local entanglement of the working substance and efficiency of the cycle. In particular cycle efficiency increases with the local entanglement $\tau_{2}$ see Fig.~\ref{Effi-Twotangle_4}

\begin{figure}[h]
\includegraphics[width=0.5\textwidth]{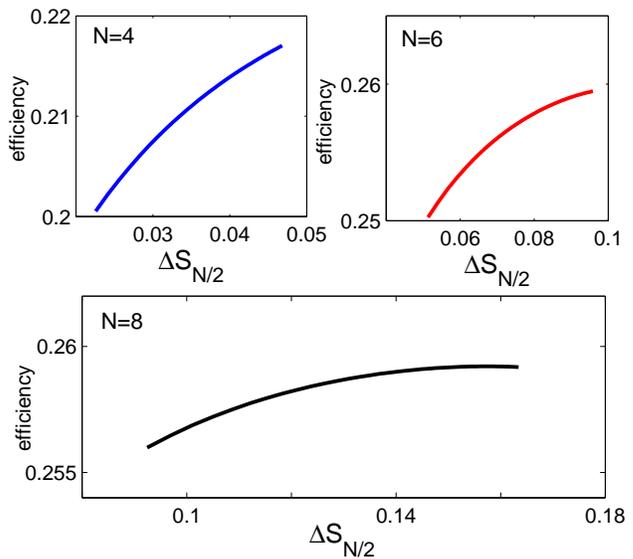}
\caption{\label{Effi-Vonneumann} The efficiency of the quantum Otto cycle as a function of the von Neumann entropy $\Delta S_{N/2}$ for three different size of the system. The parameters are $J_1=1, J_2=-1, B=0.1, d_0=2.5, d_1=1.5, T_L=10$ and $T_H=20$ to $40$.}
\end{figure}

Since local entanglement does not survive for a larger systems, in order to see connection between entanglement and cycle efficiency when increasing
the system's size we utilize von Neumann entropy (measure of the nonlocal entanglement). In particular we explore difference of the von Neumann entropy for states thermalized with the hot and cold baths respectively $\Delta S_{N/2}=S_{N/2}(T_H)-S_{N/2}(T_L)$.
For a system of $N$ spins the von Neumann entropy is defined as follows:
\begin{eqnarray}
S_{N/2}=-{\rm Tr}_{1,...,N/2}[\rho_{1,...,N/2}\log_2(\rho_{1,...,N/2})],
\label{vonneumann}
\end{eqnarray}
where, reduced density matrix for the half of the system reads $\rho_{1,...,N/2}={\rm Tr}_{N/2+1,...N}(|\Phi(t)\rangle\langle\Phi(t)|)$.
As we see Fig. \ref{Effi-Vonneumann} as large is change in the von Neumann entropy larger is the efficiency. Thus we conclude that for small system
$N=4$ engine with entangled working substance has slightly higher efficiency. For a larger systems matters difference in von Neumann entropy $\Delta S_{N/2}$ between the states thermalized with the hot and cold baths respectively.

%
%

\section{Thermal relaxation and transferred heat}
\begin{figure}
\includegraphics[width=0.45\textwidth]{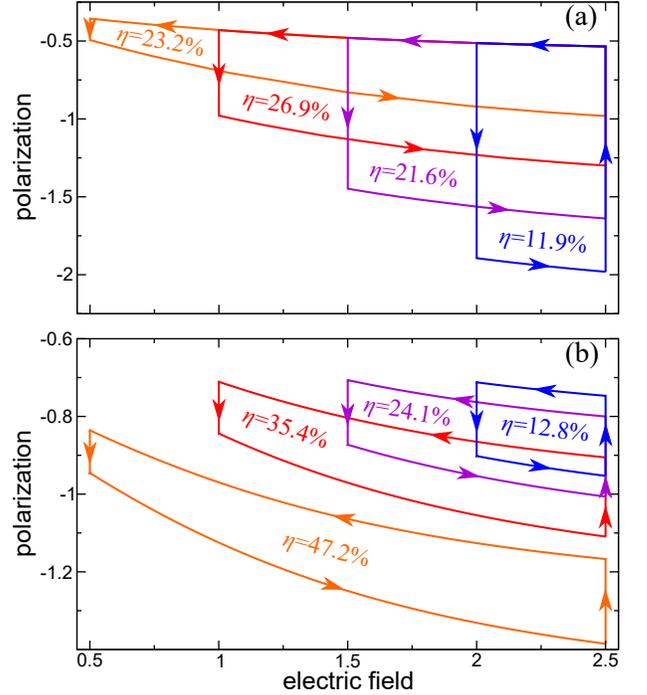}
\caption{\label{cycle} Complete quantum Otto cycle (a) using level population corresponding to Gibbs distribution and (b) level population obtained from Lindblad master equation (\ref{master equation}). The parameters are chosen as $\gamma=0.1, T_H=40, T_L=10, d_0=2.5$ and $d_1$ as in the figures.}
\end{figure}

For a derivation of the Lindblad master equation we followed  the standard procedure described in\cite{Breuer}.
We supplement the CD Hamiltonian $\hat{H}_{CD}(t)$ by the Hamiltonian of the heat bath $\hat{H}_\mathrm{bath}$ and system-bath interaction $\hat{H}_\mathrm{int}$.
In addition we assume that the phononic heat bath is coupled to the $z$ component of the vector chirality $K_n^z=(\sigma_n^x\sigma_{n+1}^y-\sigma_n^y\sigma_{n+1}^x)$. The argument behind doing this is that the vector chirality is a characteristic measure for the  non-collinearity in the spin order and is directly influenced by lattice distortion and the phononic modes
\begin{eqnarray}
\label{interactions}
&&\hat{H}=\hat{H}_{CD}(t)+\hat{H}_\mathrm{int}+\hat{H}_\mathrm{bath},\nonumber\\
&&\hat{H}_\mathrm{bath}=\int dk \omega_k \hat{b}^{\dag}_{k}\hat{b}_{k},\nonumber\\
&&\hat{H}_\mathrm{int}=\displaystyle\sum_{n=1}^4 K_n^z\int dk g_k(\hat{b}^{\dag}_{k}+\hat{b}_{k}).
\end{eqnarray}
Here $\hat{b}^{\dag}_{k},~~\hat{b}_{k}$ are the phonon creation and annihilation operators, and $g_k$ is the coupling constant between the system and the bath. After a straightforward derivations we obtain
\begin{eqnarray}
\label{master equation}
&&\frac{d\rho_S(t)}{dt}=\displaystyle\sum_{\omega,\omega^\prime}\displaystyle\sum_{\alpha,\gamma}e^{i(\omega-\omega^\prime)t}\Gamma(\omega)
\big(K_{\beta}^{z}(\omega)\rho_S(t)K_{\alpha}^{z^\dag}(\omega^\prime)\nonumber\\
&&~~~~~~~~~~~~~~~~~~~~~-K_{\alpha}^{z^\dag}(\omega^\prime)\big(K_{\beta}^{z}(\omega)\rho_S(t)\big)+h.c.,\nonumber\\
&&\Gamma(\omega)=\int_{0}^{\infty}ds e^{i\omega s}\langle B^{\dag}(t)B(t-s)\rangle.
\end{eqnarray}
Here $B(t)=\int dk g_k(\hat{b}^{\dag}_{k}e^{i\omega_{k}t}+\hat{b}_{k}e^{-i\omega_{k}t})$, $K_{\alpha}^{z}(\omega)=\displaystyle\sum_{\omega=E_m-E_n}\pi(E_n)K_{\alpha}^{z}\pi(E_n)$ and
$\pi(E_n)=\vert\Psi_n\rangle\langle\Psi_n\vert$ is the projection operator on the eigenstates $\vert\Psi_n\rangle$ of the CD Hamiltonian. For the bath correlation functions $\Gamma(\omega)$ we deduce
\begin{eqnarray}
\label{bathcorrelation}
&&\gamma(\omega)=\Gamma(\omega)+\Gamma^{\ast}(\omega),\nonumber\\
&&\gamma(\omega)=\pi J\big(\omega\big)\begin{cases} \frac{1}{\exp[\beta\omega]-1}, &  \omega< 0 \\
\frac{1}{\exp[\beta\omega]-1}+1, & \omega>0  \end{cases}.
\end{eqnarray}
Here $J\big(\omega\big)=\frac{\pi}{\omega}\displaystyle\sum_{j}g_{j}^{2}\delta(\omega-\omega_{j})=\pi\gamma$ is the spectral density of the thermal bath \cite{Breuer}. When the system relaxes the change of its energy is equal to the transferred heat. The heat absorbed by the system from the hot bath $\delta Q_{H}=(\Delta E)_H>0$ and heat released to the cold bath $\delta Q_{C}=(\Delta E)_C<0$ can be quantified in terms of the level populations $\rho_{nn}$ and energy levels $E_{n}(d)$ as follows
\begin{eqnarray}
&&(\Delta E)_H=\displaystyle\sum_{n}\rho_{nn}(d_{0},\tau_H)E_{n}(d_{0})-\displaystyle\sum_{n}\rho_{nn}(0)E_{n}(d_{0}),\nonumber\\
&&(\Delta E)_C=\displaystyle\sum_{n}\rho_{nn}(d_1,\tau_C+\tau_H)E_{n}(d_1)-\nonumber\\
&&~~~~~~~~~~~~~~~~~~~~~\displaystyle\sum_{n}\rho_{nn}(d_{0},\tau_H)E_{n}(d_{1}).
\end{eqnarray}
Here $\rho_{nn}(0)$ are the initial randomly selected level populations before contacting the system with the hot bath,
$\rho_{nn}(d_{0},\tau_H)$ are the level populations formed in the system after relaxing  to the hot bath, $\rho_{nn}(d_1,\tau_C+\tau_H)$ corresponds to the level populations formed in the system after relaxing with the cold bath, $\tau_H$ and $\tau_C$ are corresponding relaxation times. In order to recover the effect of the initial randomly selected level populations $\rho_{nn}(0)$ we run the cycle self-consistently performing several loops. Thus, we extract the values of the transferred
$\delta Q_{H}$ and released $\delta Q_{c}$ heats and estimate the efficiency of the cycle $\eta = \frac{\delta Q_{H}+\delta Q_{c}}{\delta Q_{H}}$. In order to prove that
the cycle is reversible and the spin configuration of the working substance is restored after each loop, we study the polarization of the working substance
$P=g_{ME}\displaystyle\sum_{n=1}^4(\sigma_n^x\sigma_{n+1}^y-\sigma_n^y\sigma_{n+1}^x)$ to see whether it circumscribes closed hysteresis loop.
We considered two possible scenarios of thermalization: The level populations correspond to the  Gibbs ensemble or the level
populations obtained through the Lindblad master equation. In  both cases the cycle is reversible and we observe a closed hysteresis loop
for the polarization. The thermalization time calculated via the Lindblad equation is of the order of $\tau_{th} =0.02 [\mathrm{ps}]$ smaller than
the duration of the adiabatic strokes $\tau =0.46[\mathrm{ps}]$. As we see in the Fig.~\ref{cycle}$(b)$ the working substance is restored and the cycle is reversible. The efficiency of the cycle is of the order of $\eta \approx 47\%$.

\begin{figure}
\includegraphics[width=0.95\columnwidth]{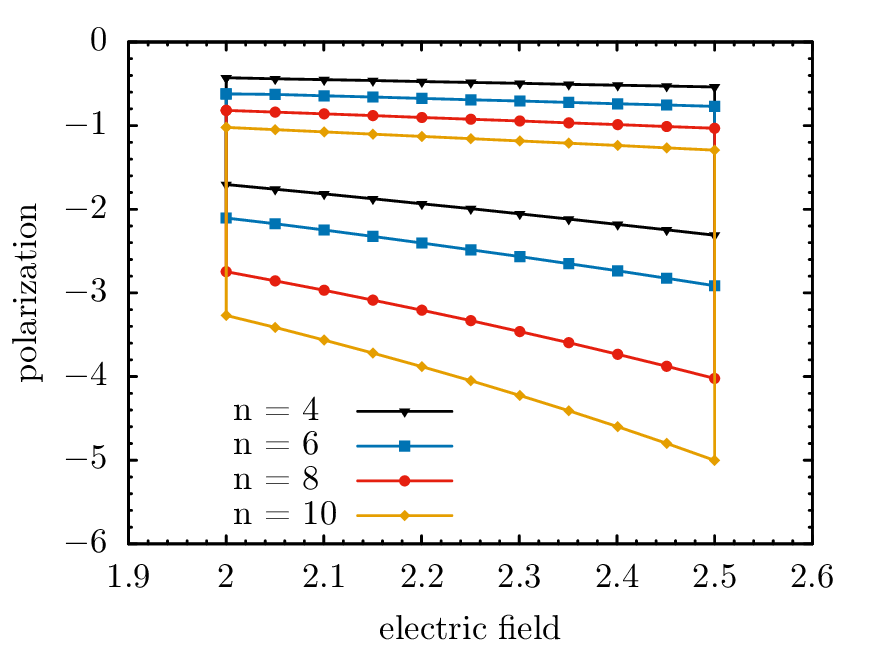}
\caption{\label{cycle_therm_eq} Complete quantum Otto cycle for different number of spins using level population corresponding to Gibbs distribution. The parameters are chosen as $\gamma=0.1, T_H=40, T_L=10, d_0=2.5$ and $d_1=2.0$.}
\end{figure}

In case of the Gibbs statistical ensemble shown in Fig.~\ref{cycle}$(a)$  we observe again that the  working substance is restored and the cycle is reversible (the polarization follows a  closed hysteresis loop). However, the efficiency is slightly different from the Lindblad case.
 The difference is an artefact of the thermalization for finite quantum systems. Footprints of the smallness and the quantumness of the working medium and even exotic properties (such as efficiency beyond the Carnot limit) were observed since the first proposals concerning the quantum heat engines \cite{Scully}. We note that in our case the  efficiency of the cycle  depends strongly  on the amplitude of the applied electric field.
The quantum Otto cycle for different number of spins is shown in Fig.~\ref{cycle_therm_eq}. As we see with increasing  the length of the spin chain the total polarization of the system becomes larger. The amplitude of variation of the polarization during the cycle enhances and the system produces more work. However, the efficiency of the cycle is the same (see Fig.~\ref{fig:effi}).

\section{Conclusion}

Our aim has been to  construct a finite time quantum thermodynamic cycle with a reasonable   output power.
Hence, we tried to minimize the execution time of the adiabatic strokes of the cycle. For this purpose we implement shortcuts to adiabaticity
and realized transitionless fast quantum adiabatic dynamics. A  merit of the present  quantum heat engine is that the working
substance experimentally feasible (e.g., single phase multiferroic $\textrm{LiCu}_{2}\textrm{O}_{2}$ spin chain) and tunable by external electromagnetic fields.
 Indeed, due to the non-collinear chiral spin order the cycle can
be controlled by an applied external electric field. For clarity we studied an exactly solvable model and
 obtained analytical expressions for the counter-diabatic Hamiltonian. Using the analytical results  the mean square fluctuation for the work, the irreversible work and output power of the cycle are evaluated.
  We observed that the work mean square fluctuations is increasing with the duration of the adiabatic strokes $\tau$  (see Fig.~\ref{rms}). However, the irreversible work shows non-monotonic behavior (see Fig.~\ref{irWork}) and has a maximum for $\tau=0.26$(ps). At the end of adiabatic stroke the irreversible work becomes zero confirming thus that the cycle is reversible. The output power of the cycle
also shows a non-monotonic behavior (see Fig.~\ref{power}) with a maximum at $\tau=0.23$(ps). This theoretical finding illustrates the existence of an inherent maximal output power. Further decreasing  the execution time of the cycle we cannot go beyond
this inherent maximal output power. By implementing a Lindblad master equation we studied the thermal relaxation of the system. We evaluated the transferred to the working substance heat $\delta Q_{H}$ and heat released by system to the cold bath $\delta Q_{c}$. We find a cylce  efficiency of  $\eta = 1+\delta Q_{c}/\delta Q_{H}\approx 47\%$. If system thermalizes to the Gibbs ensemble efficiency is lower $\eta \approx 23\%$.
\section*{Acknowledgements}
We thank Adolfo del Campo and David Zueco for the fruitful discussions. Financial support by the
Deutsche Forschungsgemeinschaft (DFG) through SFB 762,
is gratefully acknowledged.


\section{Appendix}
Eigen values $E_n$ and Eigen functions $\vert\Phi_n\rangle$ of the Hamiltonian (\ref{Hamiltonian0}), (\ref{Hamiltonian01}) in case of the four spins
\begin{eqnarray}\label{App1}
&&\vert\Phi_{1}\rangle=\vert0000\rangle,\nonumber\\
&&\vert\Phi_{2}\rangle=\frac{-i}{2}\vert1000\rangle+\frac{-1}{2}\vert0100\rangle
+\frac{i}{2}\vert0010\rangle+\frac{1}{2}\vert0001\rangle,\nonumber\\
&&\vert\Phi_{3}\rangle=\frac{i}{2}\vert1000\rangle+\frac{-1}{2}\vert0100\rangle
+\frac{-i}{2}\vert0010\rangle+\frac{1}{2}\vert0001\rangle,\nonumber\\
&&\vert\Phi_{4}\rangle=\frac{1}{2}\vert1000\rangle+\frac{-1}{2}\vert0100\rangle
+\frac{1}{2}\vert0010\rangle+\frac{-1}{2}\vert0001\rangle,\nonumber\\
&&\vert\Phi_5\rangle=\frac{1}{2}\vert1000\rangle+\frac{1}{2}\vert0100\rangle
+\frac{1}{2}\vert0010\rangle+\frac{1}{2}\vert0001\rangle,\nonumber\\
&&\vert\Phi_6\rangle=\alpha\big(\vert1100\rangle-i\mu\vert1010\rangle
-\vert1001\rangle-\vert0110\rangle\nonumber\\
&&~~~~~~~~~~~~~~~~~~~~~~~~~+i\mu\vert0101\rangle+\vert0011\rangle\big),\nonumber\\
&&\vert\Phi_7\rangle=\nu\big(\vert1100\rangle-i\lambda\vert1010\rangle
-\vert1001\rangle-\vert0110\rangle\nonumber\\
&&~~~~~~~~~~~~~~~~~~~~~~~~~+i\lambda\vert0101\rangle+\vert0011\rangle\big),\nonumber\\
&&\vert\Phi_8\rangle=\frac{1}{\sqrt{6}}\big(\vert1100\rangle+\vert1010\rangle
+\vert1001\rangle+\vert0110\rangle\nonumber\\
&&~~~~~~~~~~~~~~~~~~~~~~~~~+\vert0101\rangle+\vert0011\rangle\big),\nonumber\\
&&\vert\Phi_9\rangle=\frac{1}{\sqrt{12}}\big(\vert1100\rangle-2\vert1010\rangle
+\vert1001\rangle+\vert0110\rangle\nonumber\\
&&~~~~~~~~~~~~~~~~~~~~~~~~~-2\vert0101\rangle+\vert0011\rangle\big),\nonumber\\
&&\vert\Phi_{10}\rangle=\frac{-1}{\sqrt{2}}\vert1100\rangle
+\frac{1}{\sqrt{2}}\vert0011\rangle,~~~~~~~~~~~~~~~~~~~~~~~(A1)\nonumber\\
&&\vert\Phi_{11}\rangle=\frac{-1}{\sqrt{2}}\vert1001\rangle
+\frac{1}{\sqrt{2}}\vert0110\rangle,\nonumber\\
&&\vert\Phi_{12}\rangle=\frac{i}{2}\vert1110\rangle+\frac{-1}{2}\vert1101\rangle
+\frac{-i}{2}\vert1011\rangle+\frac{1}{2}\vert0111\rangle,\nonumber\\
&&\vert\Phi_{13}\rangle=\frac{-i}{2}\vert1110\rangle+\frac{-1}{2}\vert1101\rangle
+\frac{i}{2}\vert1011\rangle+\frac{1}{2}\vert0111\rangle,\nonumber\\
&&\vert\Phi_{14}\rangle=\frac{1}{2}\vert1110\rangle+\frac{1}{2}\vert1101\rangle
+\frac{1}{2}\vert1011\rangle+\frac{1}{2}\vert0111\rangle,\nonumber\\
&&\vert\Phi_{15}\rangle=\frac{1}{2}\vert1110\rangle+\frac{-1}{2}\vert1101\rangle
+\frac{1}{2}\vert1011\rangle+\frac{-1}{2}\vert0111\rangle,\nonumber\\
&&\vert\Phi_{16}\rangle=\vert1111\rangle,\nonumber\\
&&E_{1}=-4J_1-4J_2-4B,E_{2}=4J_2-2B-4d,\nonumber\\
&&E_{3}=4J_2-2B+4d,E_{4}=4J_1-4J_2-2B,\nonumber\\
&&E_{5}=-4J_1-4J_2-2B,\nonumber\\
&&E_6=2J_1+4J_2+2\sqrt{J_1^2+16J_2^2-8J_1J_2+8d^2},\nonumber\\
&&E_7=2J_1+4J_2-2\sqrt{J_1^2+16J_2^2-8J_1J_2+8d^2},\nonumber\\
&&E_8=-4J_1-4J_2,E_9=8J_1-4J_2,\nonumber\\
&&E_{10}=E_{11}=4J_2,\nonumber\\
&&E_{12}=4J_2+2B+4d,E_{13}=4J_2+2B-4d,\nonumber\\
&&E_{14}=-4J_1-4J_2+2B,E_{15}=4J_1-4J_2+2B,\nonumber\\
&&E_{16}=-4J_1-4J_2+4B.\nonumber
\end{eqnarray}
Here following notations are used
\begin{eqnarray}\label{App2}
&&\alpha=\frac{1}{\sqrt{4+2\mu^2}},\nonumber\\
&&\mu=\frac{4J_2-J_1
-\sqrt{J_1^2+16J_2^2-8J_1J_2+8d^2}}{2d},~~~~~~(A2)\nonumber\\
&&\nu=\frac{1}{\sqrt{4+2\lambda^2}},\nonumber\\
&&\lambda=\frac{4J_2-J_1
+\sqrt{J_1^2+16J_2^2-8J_1J_2+8d^2}}{2d}.\nonumber
\end{eqnarray}
Eigen functions of the counter-diabatic Hamiltonian $\hat{H}_{CD}(t)$
\begin{eqnarray}\label{App3}
&&\vert\Psi_{1}\rangle=\vert\Phi_{1}\rangle,\vert\Psi_{2}\rangle=\vert\Phi_{2}\rangle,
\vert\Psi_{3}\rangle=\vert\Phi_{3}\rangle\nonumber\\
&&\vert\Psi_{4}\rangle=\vert\Phi_{4}\rangle,\vert\Psi_{5}\rangle=\vert\Phi_{5}\rangle,\nonumber\\
&&\vert\Psi_{6}\rangle=C_1\big(i\frac{E_6-E_7+\sqrt{4A^2+(E_6-E_7)^2}}{2A}\vert\Phi_{6}\rangle\nonumber\\
&&~~~~~~~~~~~~~~~~~~~~~~~~~~~~~+\vert\Phi_{7}\rangle\big),\nonumber\\
&&\vert\Psi_{7}\rangle=C_2\big(i\frac{E_6-E_7-\sqrt{4A^2+(E_6-E_7)^2}}{2A}\vert\Phi_{6}\rangle\nonumber\\
&&~~~~~~~~~~~~~~~~~~~~~~~~~~~~~+\vert\Phi_{7}\rangle\big),\nonumber\\
&&\vert\Psi_{8}\rangle=\vert\Phi_{8}\rangle,\vert\Psi_{9}\rangle=\vert\Phi_{9}\rangle,~~~~~~~~~~~~~~~~~~~~~~~~~~~(A3)\nonumber\\
&&\vert\Psi_{10}\rangle=\vert\Phi_{10}\rangle,\vert\Psi_{11}\rangle=\vert\Phi_{11}\rangle,\nonumber\\
&&\vert\Psi_{12}\rangle=\vert\Phi_{12}\rangle,\vert\Psi_{13}\rangle=\vert\Phi_{13}\rangle,\nonumber\\
&&\vert\Psi_{14}\rangle=\vert\Phi_{14}\rangle,\vert\Psi_{15}\rangle=\vert\Phi_{15}\rangle\nonumber\\
&&\vert\Psi_{16}\rangle=\vert\Phi_{16}\rangle.\nonumber
\end{eqnarray}
Here
\begin{eqnarray}\label{App4}
&&C_1=\bigg\{1+\big(\frac{E_6-E_7+\sqrt{4A^2+(E_6-E_7)^2}}{2A}\big)^2\bigg\}^{-1/2},\nonumber\\
&&~~~~~~~~~~~~~~~~~~~~~~~~~~~~~~~~~~~~~~~~~~~~~~~~~~~~~~~~~~~~~~~(A4)\nonumber\\
&&C_2=\bigg\{1+\big(\frac{E_6-E_7-\sqrt{4A^2+(E_6-E_7)^2}}{2A}\big)^2\bigg\}^{-1/2}.\nonumber
\end{eqnarray}
Eigenvalues of the counter-diabatic Hamiltonian $\hat{H}_{CD}(t)$
\begin{eqnarray}\label{App5}
&&\big\{E_1,E_2,E_3,E_4,E_5,\frac{E_6+E_7+\sqrt{4A^2+(E_6-E_7)^2}}{2},\nonumber\\
&&~~\frac{E_6+E_7-\sqrt{4A^2+(E_6-E_7)^2}}{2},E_8,E_9,E_{10},~~~(A5)\nonumber\\
&&~~E_{11},E_{12},E_{13},E_{14},E_{15},E_{16}\big\}.\nonumber
\end{eqnarray}
The works produced during the quantum adiabatic strokes
\begin{eqnarray}
\label{App6}
&&\langle W_2\rangle_\mathrm{ad}=\frac{1}{Z}\bigg[\frac{2\epsilon\tau^2}{3}\bigg(e^{-\beta_{H}E_{2}(0)}-e^{-\beta_{H}E_{3}(0)}-\nonumber\\
&&~~~~~~~~~~~~~~~~~~~~~~~~~~~~~~~~~~~e^{-\beta_{H}E_{12}(0)}+e^{-\beta_{H}E_{13}(0)}\bigg)-\nonumber\\
&&~~~~~~~~~~~~~~~2\bigg(\sqrt{{J_1}^2+16{J_2}^2-8J_1J_2+8{(d_0+\frac{\epsilon\tau^2}{6})}^2}\nonumber\\
&&~~~~~~~~~~~~~~~~~~~~~~~-\sqrt{{J_1}^2+16{J_2}^2-8J_1J_2+8{d_0}^2}\bigg)\times\nonumber\\
&&~~~~~~~~~~~~~~~~~~~~\bigg(e^{-\beta_{H}E_{6}(0)}-e^{-\beta_{H}E_{7}(0)}\bigg)\bigg],~~~~~~~~(A6)\nonumber
\end{eqnarray}

\begin{eqnarray}
\label{App7}
&&\langle W_4\rangle_\mathrm{ad}=\frac{1}{Z^{\prime}}\bigg[\frac{2\epsilon\tau^2}{3}\bigg(-e^{-\beta_{L}E_{2}(\tau)}+e^{-\beta_{L}E_{3}(\tau)}+\nonumber\\
&&~~~~~~~~~~~~~~~~~~~~~~~~~~~~~~~~~~~e^{-\beta_{L}E_{12}(\tau)}-e^{-\beta_{L}E_{13}(\tau)}\bigg)-\nonumber\\
&&~~~~~~~~~~~~~~~2\bigg(\sqrt{{J_1}^2+16{J_2}^2-8J_1J_2+8{(d_0+\frac{\epsilon\tau^2}{6})}^2}\nonumber\\
&&~~~~~~~~~~~~~~~~~~~~~~~-\sqrt{{J_1}^2+16{J_2}^2-8J_1J_2+8{d_0}^2}\bigg)\times\nonumber\\
&&~~~~~~~~~~~~~~~~~~~~\bigg(-e^{-\beta_{L}E_{6}(\tau)}+e^{-\beta_{L}E_{7}(\tau)}\bigg)\bigg].~~~~~(A7)\nonumber
\end{eqnarray}

\begin{eqnarray}\label{App8}
&&\langle W_{2}^2\rangle_\mathrm{ad}=\frac{1}{Z}\bigg[\frac{4\epsilon^2\tau^4}{9}\bigg(e^{-\beta_{H}E_{2}(0)}+e^{-\beta_{H}E_{3}(0)}+\nonumber\\
&&~~~~~~~~~~~~~~~~~~~~~~~~~~~~~~~~~~~e^{-\beta_{H}E_{12}(0)}+e^{-\beta_{H}E_{13}(0)}\bigg)+\nonumber\\
&&~~~~~~~~~~~~~~~4\bigg(\sqrt{{J_1}^2+16{J_2}^2-8J_1J_2+8{(d_0+\frac{\epsilon\tau^2}{6})}^2}\nonumber\\
&&~~~~~~~~~~~~~~~~~~~~~~~-\sqrt{{J_1}^2+16{J_2}^2-8J_1J_2+8{d_0}^2}\bigg)^2\times\nonumber\\
&&~~~~~~~~~~~~~~~~~~~~\bigg(e^{-\beta_{H}E_{6}(0)}+e^{-\beta_{H}E_{7}(0)}\bigg)\bigg],~~~~~~~(A8)\nonumber
\end{eqnarray}

\begin{eqnarray}\label{App9}
&&\langle W_{4}^2\rangle_\mathrm{ad}=\frac{1}{Z^{\prime}}\bigg[\frac{4\epsilon^2\tau^4}{9}\bigg(e^{-\beta_{L}E_{2}(\tau)}+e^{-\beta_{L}E_{3}(\tau)}+\nonumber\\
&&~~~~~~~~~~~~~~~~~~~~~~~~~~~~~~~~~~~e^{-\beta_{L}E_{12}(\tau)}+e^{-\beta_{L}E_{13}(\tau)}\bigg)+\nonumber\\
&&~~~~~~~~~~~~~~~4\bigg(\sqrt{{J_1}^2+16{J_2}^2-8J_1J_2+8{(d_0+\frac{\epsilon\tau^2}{6})}^2}\nonumber\\
&&~~~~~~~~~~~~~~~~~~~~~~~+\sqrt{{J_1}^2+16{J_2}^2-8J_1J_2+8{d_0}^2}\bigg)^2\times\nonumber\\
&&~~~~~~~~~~~~~~~~~~~~\bigg(e^{-\beta_{L}E_{6}(\tau)}+e^{-\beta_{L}E_{7}(\tau)}\bigg)\bigg].~~~~~~(A9)\nonumber
\end{eqnarray}
The density matrix (\ref{master equation}) in matrix form after neglecting fast oscillating terms with $\omega=\omega^\prime$
\begin{eqnarray}\label{eqApp10}
&&\frac{d\rho_{qp}(t)}{dt}=-\frac{i}{\hbar}(E_q-E_p)\rho_{qp}+\nonumber\\
&&\displaystyle\sum_{i,j}\displaystyle\sum_{\omega_k=1}^{N}\gamma(\omega_k)\bigg[2\displaystyle\sum_{\substack{\omega_k=E_n-E_q,\\ \omega_k=E_m-E_p}}(K^z_j)_{qn}\rho_{nm}(t)(K^z_i)_{mp}\nonumber\\
&&-\displaystyle\sum_{\substack{\omega_k=E_q-E_n,\\ \omega_k=E_m-E_n}}(K^z_i)_{qn}(K^z_j)_{nm}\rho_{mp}(t)-\nonumber\\
&&\displaystyle\sum_{\substack{\omega_k=E_n-E_m,\\ \omega_k=E_p-E_m}}\rho_{qn}(t)(K^z_i)_{nm}(K^z_j)_{mp}\bigg].~~~~~~~~~~~~~~~~(A10)\nonumber
\end{eqnarray}

\begin{thebibliography}{99}

\bibitem{Landau1} L.D. Landau, E.M. Lifshitz, \textit{Quantum Mechanics: Non-Relativistic Theory}, Vol. 3, Pergamon Press, 1977.
%

\bibitem{Landau2} L.D. Landau, E.M. Lifshitz, \textit{Statistical Physics}, Vol. 5, Oxford: Pergamon Press, 1980.
%

\bibitem{Campisi} M. Campisi, P. H\"anggi, and P. Talkner, Rev. Mod. Phys. {\bf 83}, 771 (2011);
                  I. M. Sokolov, Nature Physics {\bf 10}, 7 (2014); J. P. Pekola, Nature Physics {\bf 11}, 118 (2015);
                  J. M. R. Parrondo, J. M. Horowitz, and T. Sagawa, Nature Physics {\bf 11}, 131 (2015).
%

\bibitem{Altintas} F. Altintas and \"O. E. M\"ustecaplioglu, Phys. Rev. E \textbf{92}, 022142 (2015);
                   E. A. Ivanchenko, Phys. Rev. E \textbf{92}, 032124 (2015).
%
\bibitem{Esposito2010} M. Esposito, R. Kawai, K. Lindenberg, and C. Van den Broeck, Phys. Rev. E \textbf{81}, 041106 (2010);
                       A. Alecce, F. Galve, N. Lo. Gullo, L. Dell'Anna, F. Plastina, and R. Zambrini, New J. Phys. \textbf{17}, 075007 (2015).
%

\bibitem{Georgescu} I. M. Georgescu, S. Ashhab, and F. Nori, Rev. Mod. Phys. {\bf 86}, 153 (2014);
                    H. T. Quan and F. Cucchietti, Phys. Rev. E {\bf 79}, 031101 (2009);
                    O. Abah, J. Ro{\ss}nagel, G. Jacob, S. Deffner, F. Schmidt-Kaler, K. Singer, and E. Lutz, Phys. Rev. Lett. {\bf 109}, 203006 (2012); O. Abah and E. Lutz, EPL {\bf 106}, 20001 (2014); S. \c Cakmak, F. Altintas, \"O. E. M\"ustecaplioglu, arXiv:1510.04495.
%

\bibitem{Rossnagel} J. Ro{\ss}nagel, O. Abah, F. Schmidt-Kaler, K. Singer, and E. Lutz, Phys. Rev. Lett. {\bf 112}, 03602 (2014);
                    A. M. Zagoskin, S. Savelev, F. Nori, and F. V. Kusmartsev, Phys. Rev. B {\bf 86}, 014501 (2012);
                    N. Linden, S. Popescu, and P. Skrzypczyk, Phys. Rev. Lett. {\bf 105}, 130401 (2010);
                    R. Wang, J. Wang, J. He, and Y. Ma, Phys. Rev. E {\bf 86}, 021133 (2012);
                    A. \"U. C. Hardal and \"O. E. M\"ustecaplioglu, Scientific Reports \textbf{5}, 12953 (2015).
%

\bibitem{Jarzynski1} C. Jarzynski, Phys. Rev. A {\bf 88}, 040101(R) (2013);
                     A. del Campo, Phys. Rev. Lett. {\bf 111}, 100502 (2013).

%

\bibitem{delCampo2014} A. del Campo, J. Goold, and M. Paternostro, Sci. Rep. {\bf 4}, 6208 (2014).

\bibitem{delCampo} A. del Campo, M. M. Rams, W. H. Zurek, Phys. Rev. Lett. {\bf 109}, 115703 (2012);
                   H. Saberi, T. Opatrn\'y, K. M\o lmer, and A. del Campo, Phys. Rev. A {\bf 90}, 060301(R) (2014).
%

\bibitem{Wang2} H. Wang and G. Wu, Phys. Lett. A {\bf 376}, 2209 (2012);
                S. N. Shevchenko, D. G. Rubanov, and Franco Nori, Phys. Rev. B \textbf{91}, 165422 (2015).
%
\bibitem{Esposito} M. Esposito, N. Kumar, K. Lindenberg, and C. Van den Broeck, Phys. Rev. E {\bf 85}, 031117 (2012);
                   N. Kumar, C. Van den Broeck, M. Esposito, and K. Lindenberg, Phys. Rev. E {\bf 84}, 051134 (2011).
%

\bibitem{Poletti} Y. Zheng, S. Campbell, G. De Chiara, and D. Poletti, arXiv:1509.01882v2; Y. Zheng, D. Poletti
Phys. Rev. E \textbf{92}, 012110 (2015); Y. Zheng, D. Poletti, Phys. Rev. E \textbf{90}, 012145 (2014).

%
\bibitem{Demi} M. Demirplak and S. A. Rice, J. Phys. Chem. A \textbf{107}, 9937 (2003), J. Phys. Chem. B \textbf{109}, 6838 (2005).

%
\bibitem{Berry} M. V. Berry, J. Phys. A: Math. Theor. {\bf 42}, 365303 (2009).

%
\bibitem{Azimi1} M. Azimi, L. Chotorlishvili, S. K. Mishra, S. Greschner, T. Vekua, and J. Berakdar, Phys. Rev. B {\bf 89}, 024424 (2014);
                 L. Chotorlishvili, R. Khomeriki, A. Sukhov, S. Ruffo, and J. Berakdar, Phys. Rev. Lett. {\bf 111}, 117202 (2013).
%
\bibitem{Azimi2} M. Azimi, L. Chotorlishvili, S. K. Mishra, T. Vekua, W. H\"ubner, and J. Berakdar, New J. of Phys. {\bf 16}, 063018 (2014).
%

\bibitem{Wang3} J. Wang \textit{et al.}, Science {\bf 299}, 1719 (2003);
                W. Eerenstein, N. D. Mathur, and J. F. Scott, Nature {\bf 442}, 759 (2006);
                V. Garcia \textit{et al.}, Science {\bf 327}, 1106 (2010).
%

\bibitem{Bibes} M. Bibes and A. Barthelemy, Nat. Mater. {\bf 7}, 425 (2008);
                S. W. Cheong and M. Mostovoy, Nat. Mater. {\bf 6}, 13 (2007).
%

\bibitem{Dawber} M. Dawber, K. M. Rube, and J. F. Scott,  Rev. Mod. Phys. {\bf 77}, 1083 (2005);
                 C. G. Duan, S. S. Jaswal, and E. Y. Tsymbal, Phys. Rev. Lett. {\bf 97}, 047201 (2006).
%

\bibitem{Valencia} S. Valencia \textit{et al.}, Nat. Mater. {\bf 10}, 753 (2011).
%


\bibitem{Sahoo} S. Sahoo, S. Polisetty, C. G. Duan, S. S. Jaswal, E. Y. Tsymbal, and C. Binek, Phys. Rev. B {\bf 76}, 092108 (2007);
                N. Kida and Y. Tokura, J. Magn. Magn. Mater. {\bf 324}, 3512 (2012).
%

\bibitem{Menzel} M. Menzel \textit{et al.}, Phys. Rev. Lett. {\bf 108}, 197204 (2012);
                 A. J. Hearmon \textit{et al.}, Phys. Rev. Lett. {\bf 108}, 237201 (2012).
%
\bibitem{Mostovoy} M.  Mostovoy, Phys. Rev. Lett. {\bf 96}, 067601 (2006);
                   H.  Katsura, N. Nagaosa, and A. V. Balatsky, Phys. Rev. Lett. {\bf 95}, 057205 (2005).
%

\bibitem{rep_prog_phys}
Y. Tokura and S. Seki, Advanced materials \textbf{22}, 1554-1565 (2010); 
Y. Tokura, S. Seki, and N. Nagaosa, Rep. Prog. Phys. \textbf{77}, 076501 (2014).

%
\bibitem{fiebig_science} N.A. Spaldin and M. Fiebig, Science \textbf{309}, 391 (2005).

%
\bibitem{Park} S.  Park, Y. J. Choi, C. L. Zhang, and S. W. Cheong, Phys. Rev. Lett. {\bf 98}, 057601 (2007);
               Y. Yamasaki, S. Miyasaka, Y. Kaneko, J.-P. He, T. Arima, and Y. Tokura, Phys. Rev. Lett. \textbf{96}, 207204 (2006).
               Y. Yasui, Y. Yanagisawa, R. Okazaki, and I. Terasaki, Phys. Rev. B \textbf{87}, 054411 (2013).
%
\bibitem{Loidl}
F. Schrettle, S. Krohns, P. Lunkenheimer, J. Hemberger, N. B\"uttgen, H.-A. Krug von Nidda,
               A. V. Prokofiev, and A. Loidl, Phys. Rev. B \textbf{77}, 144101 (2008).

%
\bibitem{wiesendanger}
A. A. Khajetoorians, M. Steinbrecher, M. Ternes, M. Bouhassoune, M. dos Santos Dias, S. Lounis, J. Wiebe, and R. Wiesendanger,
Nature Communications \textbf{7}, 10620 (2016).
%

\bibitem{torron}   E. Torrontegui, S. Ibanez, S. Martinez-Garaot, M. Modugno, A. del Campo, D. Guery-Odelin, A. Ruschhaupt, X. Chen, and J. G. Muga,
                   Advances in Atomic, Molecular, and Optical Physics, \textbf{62}, 117-169 (2013).

%
\bibitem{Scully}
M. O. Scully, M. S. Zubairy, G. S. Agarwal, and H. Walther, Science \textbf{299}, 862 (2003).
%
\bibitem{Bochkov}
G. N. Bochkov and Yu. E. Kuzovlev, Sov. Phys. JETP \textbf{45}, 125 (1977)
%
\bibitem{Kuzovlev}
G. N. Bochkov and Yu. E. Kuzovlev, Physics Uspekhi \textbf{56}, 590 (2013).
%
\bibitem{Jarzynski2} C. Jarzynski, Phys. Rev. Lett. {\bf 78}, 2690 (1997).
%
\bibitem{Deffner}
S. Deffner and E. Lutz, Phys. Rev. Lett \textbf{105}, 170402 (2010);
P. Talkner, E. Lutz, and P. H\"anggi, Phys. Rev. E \textbf{75}, 050102(R) (2007).
%

\bibitem{Amico}  L. Amico, R. Fazio, A. Osterloh, and V. Vedral, Rev. Mod. Phys. \textbf{80}, 517 (2008);
                 W. K. Wootters, Phys. Rev. Lett. \textbf{80}, 2245 (1998).
%
\bibitem{ChenBo}
H.-B. Chen, Y.-Q. Li, and J. Berakdar, J. Appl. Phys. \textbf{117},  043910 (2015);
J.-H. Moon, S.-M. Seo, K.-J. Lee, K.-W. Kim, J. Ryu, H.-W. Lee, R. D. McMichael, and M. D. Stiles, Phys. Rev. B \textbf{88}, 184404 (2013).
%
\bibitem{Breuer} H. P. Breuer and F. Petruccione, \textit{The Theory of Open Quantum Systems}, Oxford University Press, Oxford, (2002).
%

\end{thebibliography}
\end{document}